# A Novel CustNetGC Boosted Model with Spectral Features for Parkinson's Disease Prediction

Abishek Karthik[1], Pandiyaraju V[2], Dominic Savio M[3], Rohit Swaminathan S[4]

***Abstract***

Parkinson's disease is a neurodegenerative disorder that can be very tricky to diagnose and treat. Such early symptoms can include tremors, wheezy breathing, and changes in voice quality as critical indicators of neural damage. Notably, there has been growing interest in utilizing changes in vocal attributes as markers for the detection of PD early on. Based on this understanding, the present paper was designed to focus on the acoustic feature analysis based on voice recordings of patients diagnosed with PD and healthy controls (HC). Previous studies have indicated that PD affects most aspects of speech, underlining the differences between affected and unaffected individuals.

In this paper, we introduce a novel classification and visualization model known as CustNetGC, combining a Convolutional Neural Network (CNN) with Custom Network Grad-CAM and CatBoost to enhance the efficiency of PD diagnosis. We use a publicly available dataset from Figshare, including voice recordings of 81 participants: 40 patients with PD and 41 healthy controls. From these recordings, we extracted the key spectral features: L-mHP and Spectral Slopes. The L-mHP feature combines three spectrogram representations: Log-Mel spectrogram, harmonic spectrogram, and percussive spectrogram, which are derived using Harmonic-Percussive Source Separation (HPSS). Grad-CAM was used to highlight the important regions in the data, thus making the PD predictions interpretable and effective.

Our proposed CustNetGC model achieved an accuracy of 99.06% and precision of 95.83%, with the area under the ROC curve (AUC) recorded at 0.90 for the PD class and 0.89 for the HC class. Additionally, the combination of CatBoost, a gradient boosting algorithm, enhanced the robustness and the prediction performance by properly classifying PD and non-PD samples. Therefore, the results provide the potential improvement in the CustNetGC system in enhancing diagnostic accuracy and the interpretability of the Parkinson's Disease prediction model.

## 1. INTRODUCTION

Parkinson's disease is a complex, multi-faceted neurodegenerative disorder that causes impairment in various motor and non-motor functions of the nervous system. It is characterized by gradual degeneration in the CNS generally associated with impairment of motor functions, making it challenging to diagnose [2]. As of 2016, more than 6.06 million people in the world have PD, after Alzheimer's disease. Besides this, the continuous progression of the disease devastates the quality of life for most patients as they lose both physical and cognitive capabilities and thus result in fatalities. Reports claim that PD contributed to 211.3 thousand deaths in 2016 [3].

Although there isn't a decisive remedy for the treatment of PD, early as well as the exact diagnosis really contribute to progression with timely interventions coupled with appropriate specific treatment. Primary signs of PD involve subliminal, such as a tremble or shivering body and voice variation.

The timing as well as right detection is absolutely fundamental in initiation of proper medicare and application of specific remedy treatment.

There's one marked area of PD-specific difference: Its interference with one's speech pattern. Individuals with PD tend to show hypophonia, a very soft and sometimes inaudible voice due to the degeneration of neural circuits involved in regulating speech-producing muscles. These vocal manifestations are one of the most informative markers for diagnosis and could be analyzed rather efficiently using techniques in advanced deep learning. Specifically, L-mHP and Spectral Slopes features can be exploited to identify PD from speech signals. They include discrete spectral characteristics and provide an effective representation of audio signals. The log-mel spectrogram captures the global acoustic properties, whereas the harmonic spectrogram focuses on the frequency distribution and the percussive spectrogram focuses on transient features. Thus, the combination of these features presents a solid basis for PD detection.

This research study explores feature extraction of L-mHP from recordings of voices with the aid of Spectral Slopes for differentiating between healthy subjects and PD-afflicted individuals. The features have been further extracted and then filtered through a deep customized learning framework designed with the application of an Xception fine-tuned, enhanced CatBoost model with gradients. The proposed model was trained, validated, and tested on the Figshare dataset, which holds voice recordings of individuals with and without PD [5]. Those performance metrics were applied, such as accuracy, precision, and recall to measure the predictive power of the model. MFCC and HFCC also were applied for feature extraction, and optimization of parameters is one of the major challenges while building the proposed CustNetGC model.

The advent of deep learning, especially Convolutional Neural Networks (CNNs), has revolutionized computational methods in biological research. Unlike traditional machine learning classifiers, which mainly rely on handcrafted global features (e.g., pitch, jitter, shimmer, and MFCC), CNNs are more adept at extracting hierarchical and localized features, making them well-suited for complex classification tasks. Despite their computational demands, CNN ensembles, leveraging checkpoint-based or snapshot-based techniques, can significantly enhance predictive accuracy. This paper also discusses several fine-tuning strategies to make the training of CNNs easier and more efficient.

## 2. LITERATURE SURVEY

Karaman et al. (2021) has conducted much research on the early and precise diagnosis of Parkinson's Disease (PD). They applied CNN architectures, including ResNet101, DenseNet161, and SqueezeNet1_1, for fine-tuning the mPower Voice database. In this experiment, DenseNet161 performed with a high accuracy of 89.75%, precision of 88.40%, and sensitivity of 91.50%. In addition, Iyer et al. (2023) implemented the use of Inception V3 with transfer learning to inspect voice spectrograms, and it was able to give insights into class-specific area under the curve values. The performances of the machine learning models of SVM, Naïve Bayes, KNN, and ANN are compared by Rana et al. (2023), with a maximum accuracy of 96.7% given by the model of ANN. Xu et al. (2020) proposed Spectrogram-Deep Convolutional GAN to augment the data while using ResNet50 and Global Average Pooling in order to enhance feature extraction and classify voiceprint data. Majda-Zdancewicz et al. (2021) used AlexNet and SVM, with sensitivity of 97% and compared conventional signal classification approaches. Quan et al. (2021) used LSTM algorithms with moderate accuracy of 73.35%. The accuracies of the SVM, Random Forest, and Logistic Regression model attained in Govindu et al. (2023) are 91.83%, and sensitivity with SVM is around 0.95. Here also, those architectures use pre-trained versions. Senturk et al. (2020) have reported Regression Trees, Neural Networks, and SVM in its usage. And they have claimed that accuracy to reach 93.84%. Besides, some newly evolved detection methods in the literature report. Khosla et al. (2024) applied Freezing-of-Gait (FOG) detection with impressive metrics such as 99% accuracy, 97.4% precision, and 99.1% sensitivity.

Borzì et al. (2020) applied a FOG detection algorithm with minimal data and reported achieving 85.5% accuracy using leaveone- subject-out validation. Goyal et al. (2021) discussed the application of hybrid CNN models in vocal loss detection with 99.37% accuracy. Zahid et al. (2020) focused on the speech spectrograms and achieved a diagnostic accuracy of 99% with machine learning classifiers. Hireš et al. (2021) used CNNs and ROC curve analysis for the detection of PD based on voice recordings from 50 patients with PD and 50 healthy subjects. Faragó et al. (2023) evaluated noise reduction techniques applying Wiener filters on continuous speech recordings in noisy environments for improved differentiation of Parkinsonian speech. Guatelli et al. (2023) experimented with Extreme Learning Machines and CNN models based on the spectrogram, which were the best detection accuracy by ResNet50. Senturk et al. (2022) reported 99.74% accuracy using RNNs and CFNNs for voice-based PD detection, while Er et al. (2021) employed LSTM models integrated with ResNet-18, ResNet-50, and ResNet-101, achieving reliable classification from mel-spectrograms.

Celik et al. (2023) proposed a SkipConNet + RF model with 99.11% accuracy via a CNN-based approach and diagnosis using a unique method. Narasimha Rao et al. (2023) proposed a three-stage classification framework using Optimized ResNet, GoogleNet, and RBF-Gated Recurrent Unit named ORG-RGRU. Nijhawan et al. (2023) compared the best state-of-the-art models via MLP, SVM, Random Forest, and Gradient-Boosted Decision Trees. They proved that the best model is Gradient Boosting. Several studies researched feature engineering in improving the detection accuracy. Hawi et al. (2022) demonstrated that the long-term features combined with MFCCs improved the model performance. Solana-Lavalle et al. (2020) attained 94.7% accuracy, 98.4% sensitivity, and 97.22% precision with MLP, SVM, and RF. Zhang et al. (2021) used energy direction-based features achieved by using Empirical Mode Decomposition and effectively distinguished PD from healthy candidates.

## Table 1. Summary of Literature Survey

| Techniques | Performance | Demerits |
|---|---|---|
| SqueezeNet1_1, ResNet101, DenseNet161 | Accuracy: 89%, Sensitivity: 91.5%, Precision: 88.4% | Accuracy < 90% |
| Inception V3 with Transfer Learning | High AUC | Difficulty identifying spectrogram regions |
| SVM, NaÃ¯ve Bayes, KNN | Accuracy: SVM: 87.17%, NaÃ¯ve Bayes: 74.11%, KNN: 87.17% | Low accuracy |
| S-DCGAN | Accuracy: 91.25% | Limited patient voiceprint datasets |
| SVM, AlexNet CNN | Sensitivity: 97% | Data inaccuracies from real recordings |
| LSTM | Accuracy: 73.35%, F-score: 79.67%, MCC: 0.3773 | Accuracy < 90% |
| SVM, Random Forest, KNN, Logistic Regression | Accuracy: 91.83%, Sensitivity: 95% | Reliance on pre-trained models |
| Regression Trees, Neural Networks, SVM | Accuracy: 93.84% | Reliance on pre-trained models |
| Freezing-of-Gait (FoG) Detection | Precision: 99%, Sensitivity: 97.4%, Accuracy: 99.1%, F1-score: 98.1% | Reliance on pre-trained models |
| CNN Hybrid Method | Accuracy: 99.37% | Limited data used |
| ResNet50 | Superior performance to CNN | Lack of specific region selection |
| Logistic Regression, Random Forest, Gradient Boosted Trees | Recall: 0.797, Precision: 0.901, F1: 0.836 | Reliance on pre-trained models |
| Transfer Learning Using Speech Spectrograms | Accuracy: 99.1% | Limited data, reliance on pre-trained models |
| CNN for PD Detection from Voice | Sensitivity: 86.2%, Specificity: 93.3%, AUC: 89.6% | Lack of specific region selection |
| CFNN, RNN, Feedforward NNs | RNN with 300 voice features | Lack of specific region selection |
| Speech Classification Using CNN | Accuracy: Speech: 93%, Energy: 96%, Mel Spectrogram: 92% | Noisy environments |
| AlexNet, VGG-16, SqueezeNet, Inception V3, ResNet-50 | High accuracy from pre-trained CNNs | Long training times |
| LSTM | High classification accuracy | Lack of specific region selection |
| SkipConNet + RF | Accuracy: 99.11% | Evaluation on pre-trained model |
| MFCC + Long-Term Features | Accuracy: 88.84% | Accuracy < 90% |
| Optimized ResNet, GoogleNet, RGRU | Accuracy: 95%, MCC: 91% | Use of traditional approaches |
| GBDTs vs. MLP, SVM, RF | GBDTs outperform others by 1% AUC | Conventional ML techniques used |
| k-NN, MLP, SVM, RF | Accuracy: 94.7%, Sensitivity: 98.4%, Specificity: 92.68%, Precision: 97.22% | Reliance on pre-trained models |
| EDF-EMD Features | Accuracy: Sakar: 96.54%, CPPDD: 92.59% | No DL used, reliance on pre-trained models |

## 3.PROPOSED SYSTEM ARCHITECTURE

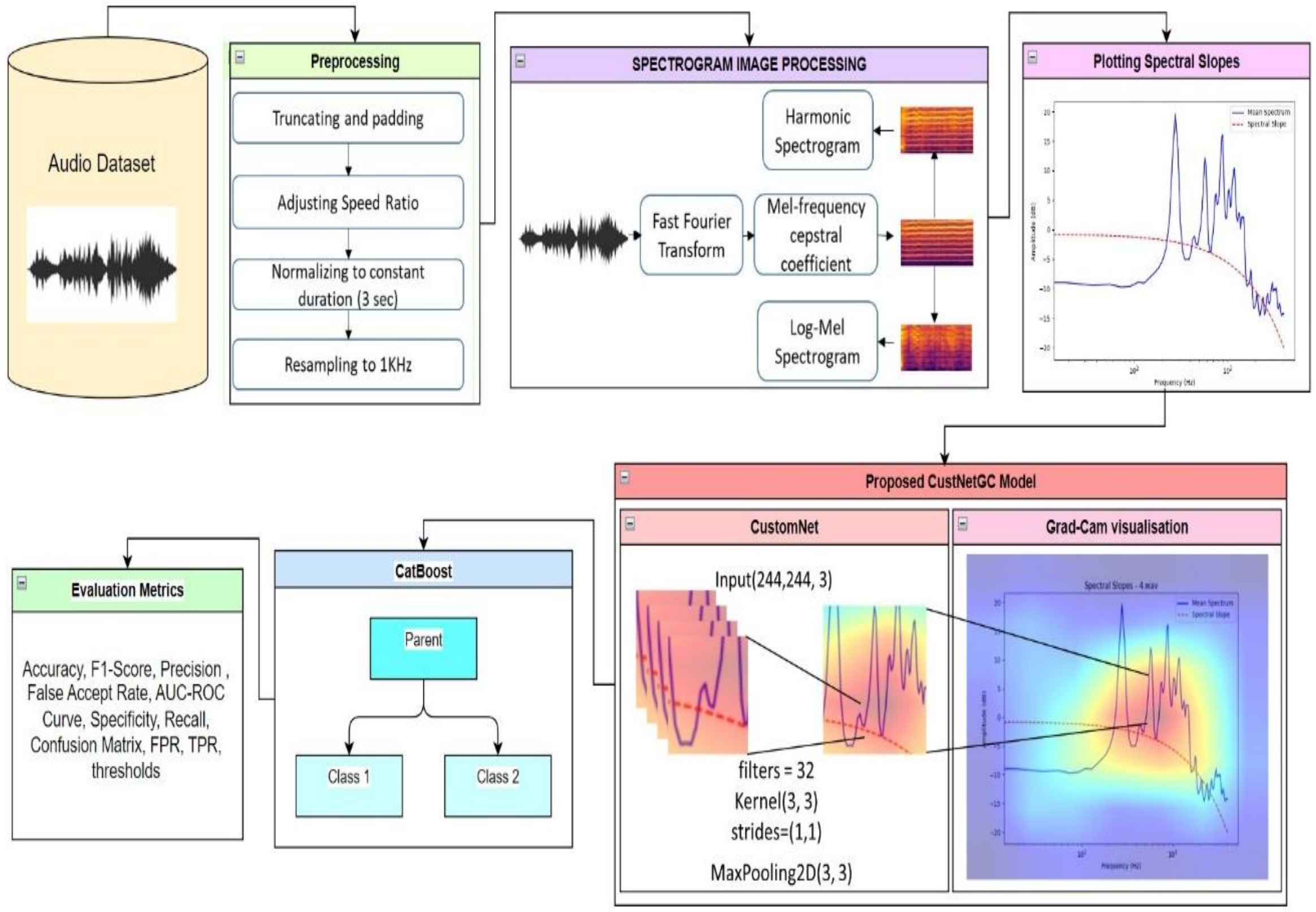

**Figure 1. Proposed CustNetGC Boosted model**

The audio dataset utilized in this experiment consists of 81 voice recordings that are from both patients with Parkinson's disease (PD) and healthy controls (HC). Since the recordings varied widely in their duration, their spectrogram lengths were inconsistent. A speed ratio formula was utilized to standardize the duration across the dataset, and then the adjusted audio samples were resampled. These voice samples were then converted into Log-Mel spectrograms, harmonic spectrograms, and percussive spectrograms. Combined spectrograms will give an inclusive visual representation of the voice records carrying additional information from each description.

Clearly, the differentiation pattern between the audio samples from patients with PD versus those without it is reflected clearly in the plots of spectral slopes. These patterns were fed into a custom-trained Xception model, which is a CNN optimized for classification tasks involving Grad-CAM images. Finally, the model's predictions were refined using CatBoost, a gradient boosting algorithm, to enhance classification accuracy. The proposed system has been verified based on measures such as Accuracy, F1-Score, Precision, FAR, AUC-ROC Curve, Specificity, Recall, Confusion Matrix, False Positive Rate, True Positive Rate, and Thresholds.

### 3.1 Dataset Exploration

The audio dataset was retrieved from the publicly available Figshare repository. Voice recordings are available in the format of.wav files. They contain recordings where the vowel /a/ is pronounced very slowly. Recordings were conducted by using the phones of the participants. A total of 81 voices of 40 individuals with PD and 41 HC voices are used as shown in Figure 2. No history of Parkinson's disease, Parkinsonian symptoms, or other neurological/psychological disorder has been found among the HC participants.

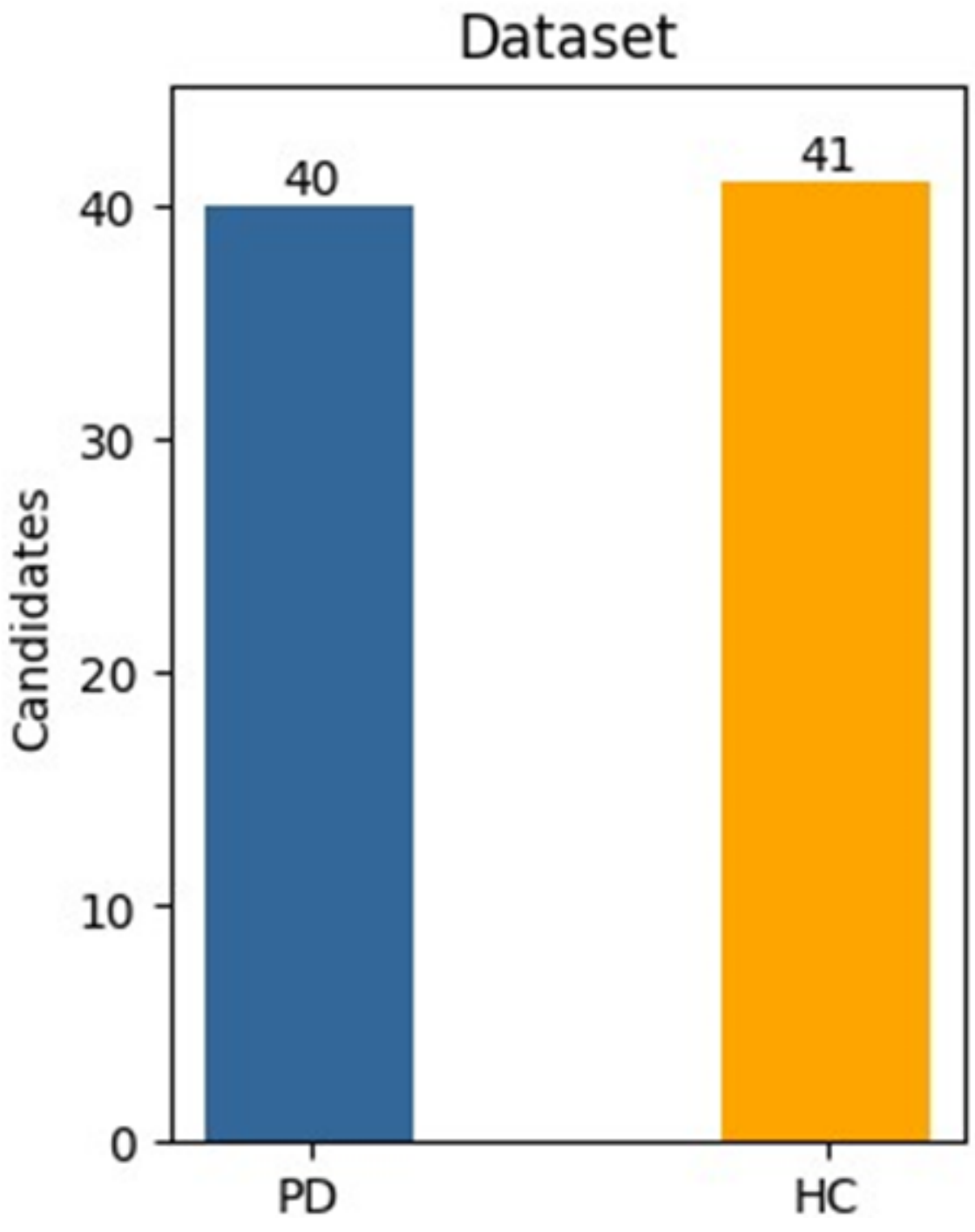

**Figure 2. Dataset Distribution Graph**

This audio dataset contains voices of Parkinson. In paper [2] the type of connection between attributes at higher and lower structural levels is described. However, this is there is still a difficult in studying on the English phonetics, and the reader will undoubtedly appreciate the author's clear vision and strong sense of purpose — and possibly gain some new insights as well as been mentioned. In 2023, Iyer, Anu, et al used the ML method to process voice samples for identification of PD [5] in which they had used this same dataset for the work.

**Table 2. Sample of Dataset Demographics**

| Sample ID | Label | Age | Sex |
|---|---|---|---|
| AH_064F_7AB034C9-72E4-438B-A9B3-AD7FDA1596C5 | HC | 69 | M |
| AH_114S_A89F3548-0B61-4770-B800-2E26AB3908B6 | HC | 43 | M |
| AH_121A_BD5BA248-E807-4CB9-8B53-47E7FFE5F8E2 | HC | 18 | F |
| AH_123G_559F0706-2238-447C-BA39-DB5933BA619D | HC | 28 | M |
| AH_195B_39DA6A45-F4CC-492A-80D4-FB79049ACC22 | HC | 68 | M |
| AH_197T_7552379A-2310-46E1-9466-9D8045C990B8 | HC | 24 | M |
| AH_545622717-461DFFFE-54AF-42AF-BA78-528BD505D624 | PwPD | 77 | M |
| AH_545622718-C052AD58-5E6B-4ADC-855C-F76B66BAFA6E | PwPD | 72 | F |
| AH_545622719-52C23861-6E0D-41E0-A3D8-9358C28C019B | PwPD | 69 | F |
| AH_545622720-E1486AF6-8C95-47EB-829B-4D62698C987A | PwPD | 68 | M |
| AH_545622722-3C79DA68-36BB-43A2-B29C-61AEF480E07E | PwPD | 69 | M |
| AH_545629296-C2C009C6-8C17-42EA-B6BE-362942FC4692 | PwPD | 74 | F |

These samples are total of 81 voice recordings in which 40 are of Parkinson Diseased persons and the rest 41 are the voice of healthy candidates. Some samples of dataset is provided in table 2. These audio recordings are of different persons are of uneven durations of length in the recording. Since these recordings have frequently varied lengths in their durations it is impossible to convert them into batch inputs to the machine learning models. But there are different ways to pre-processing the data.

### 3.2 Preprocessing

To balance the issue with unequal audio time lengths in the dataset, there could be some utilization of making tempo adjustments on voice samples. With playback speed, the alteration involves both changing durations and pitches because the tempo manipulation modifies the record frequency proportionate to the actual playback speed at which the playback takes place. We thus use a greater sample rate for the original audio in slowing down the playback to give quality audio since a low playback rate can reduce the audio clarity. The process is repeated for all audio samples in the dataset and gives uniformity in target duration.

$$speed_ratio = \frac{target_duration}{current_duration} \quad \text{------- (1)}$$

Take the equation (1) and multiplied with the playback speed we get the adjusted audio length.

$$adjusted_audio = playback_speed \; X \; speed_ratio \quad \text{------- (2)}$$

Standardization of the audio sample length can make it possible to generate spectrograms of voice recordings with equal lengths for effective feature analysis. Previous researches [4] have found CNN architectures and feature selection techniques that improve the speech pattern-based classification of PD cases. Altering the playback speed may reduce audio quality, hence, voice characteristic analysis may become complicated. Calculating voice characteristics may then be wrong. To counter this, other preprocessing techniques, including padding and truncation, are applied to normalize the duration of the audio recordings effectively.

Padding and truncation help to ensure uniformity in the lengths of audio files. Padding adds zeros to shorter sequences, making them equal in length to the longest sample in the dataset or to the maximum allowable length of the model. On the other hand, truncation shortens longer sequences to the desired length. This way, audio samples are turned into rectangular tensors suitable for machine learning models. Padding and truncation is performed such that all samples become uniform based on some given equations. When the process of padding and truncation is implemented, the audio samples are normalized to three seconds in length. Following the above two procedures, normalizing the dataset, with a purpose of taming noise and scales. The purpose of normalization is normalization so as to maintain the data with a uniform range by making proper scaling and adjustments within the signal. This applies a constant gain, which may be either positive or negative, to all recordings, meaning that all samples in the dataset would have similar levels of amplitude. Without normalizing, errors in audio scale might result in misinterpretations of signals; hence, feature extraction will be compromised.

Truncation,

$$truncated_sample = round\left\{\frac{original_sample}{scaling_factor}\right\}$$

------- (3)

From equation (3) we get the value for the truncated sample.

For Padding,

$$num_zeros_to_add = desired_length - len(original_samples)$$

------- (4)

From equation (4) we got the value for the number of zeros that needed to be added for original samples then it is used in the equation (5) to get the padded signal.

$$padded_signal = original_samples + [0] * num_zeros_to_add$$

------- (5)

Normalization further normalizes volume levels across the dataset. Sample rate is a critical parameter in digital audio processing that refers to the number of samples taken per second and is measured in Hertz (Hz). Calculating and maintaining the right sample rate is important to ensure that audio data does not get deteriorated. The formula to compute sample rate is as follows:. This step prepares the dataset for further robust and accurate analysis in the subsequent stages:

$$Sample\ Rate\ (SR) = 2\ X\ F_{max}$$

------- (6)

Where:

- $F_{max}$ – Represents the maximum frequency recorded in the audio signal.

Sample rate is calculated by 2 times maximum frequency recorded in the audio signal but we alternatively using the number of samples captured,

$$Sample\ Rate\ (SR) = \frac{N_s}{T}$$

------- (7)

Where:

- $N_s$ – Represents the number of samples captured.
- T – Represents the duration of the audio in seconds.

Here T = 3 as we altered the length durations of the audio dataset to 3 seconds. After computations we found the value of $N_s$ to be 24000. That is 24 thousand of samples captured in duration of 3 seconds. So, the calculated sample rate is 8000 that is 8 thousand of samples captured per second. For normalization we need to change the sample rate of the audio data.

The sample rate modification process ensures that all voice samples are of the same format, which makes their processing easier. After repeated normalization, it was found that resampling the audio samples to a rate of 1 kHz made the accuracy of the feature extraction phase better. So, all the audio samples are resampled to 1 kHz. The processed audio samples are then stored in a specific file, so that it becomes easy to handle them for the next steps.

Next, Log-Mel spectrograms are computed for the processed audio samples and saved individually. Spectrograms are graphical representations of the frequency content of sound waves recorded over time and are rapidly gaining popularity in voice signal analysis. In essence, a spectrogram is the squared magnitude spectrum of an audio signal. Using Log-Mel spectrograms, the deep learning model uses well-established techniques in image classification. These spectrograms feed the model in a manner that is almost like the input to human senses, thereby enhancing the capacity of the system to extract meaningful features from the data.

The passing of unprocessed audio waves through filter banks generates a Log-Mel spectrogram. This is achieved by using a frequency-domain filter bank on windowed audio streams. The centre frequencies of the filters, and consequently their corresponding time instants for the analysis windows, are the two- and three-dimensional representations of the second and third outputs of the spectrogram, respectively. By definition of the Log-Mel scale, intervals with identical sound should give consistent measurements. This property makes Log-Mel spectrograms well-suited for the task of segmenting signals into uniform periods in a wide array of audio processing applications.

The bands in the Log-Mel spectrogram can be divided based on either the Log-Mel scale or the Hertz scale. The former is relatively preferred as it is closer to the human interpretation of acoustic perception, therefore permitting a more intuitive representation of sound frequencies. This also makes it a very good tool for deep learning models. The steps and equations for computing Log-Mel spectrograms are as follow.

The equation from O'Shaughnessy 1987 [39] formula to find *M* Mel from *f* Hertz is,

$$M = 2595\left(1 + \frac{f}{700}\right)$$

------- (8)

Now,

$$M_n(f) = \sum_{f'=0}^{N-1} X\,(f', t).\,H_n(f')$$

------- (9)

Where,

X (f, t) – Represents the magnitude of the spectrum obtained,

f – Represents the frequency index and,

t – Represents the time index.

$H_n(f)$ – Represents the nth Mel filter.

The Logarithmic Transformation is:

$$L_n(t) = log\ log\ \left(M_n(t)\right)$$

------- (10)

Here,

$M_n(t)$ – Represents the nth Mel spectrogram frame at time index t and,

$L_n(t)$ – Represents the nth Log – Mel spectrogram frame at time index t.

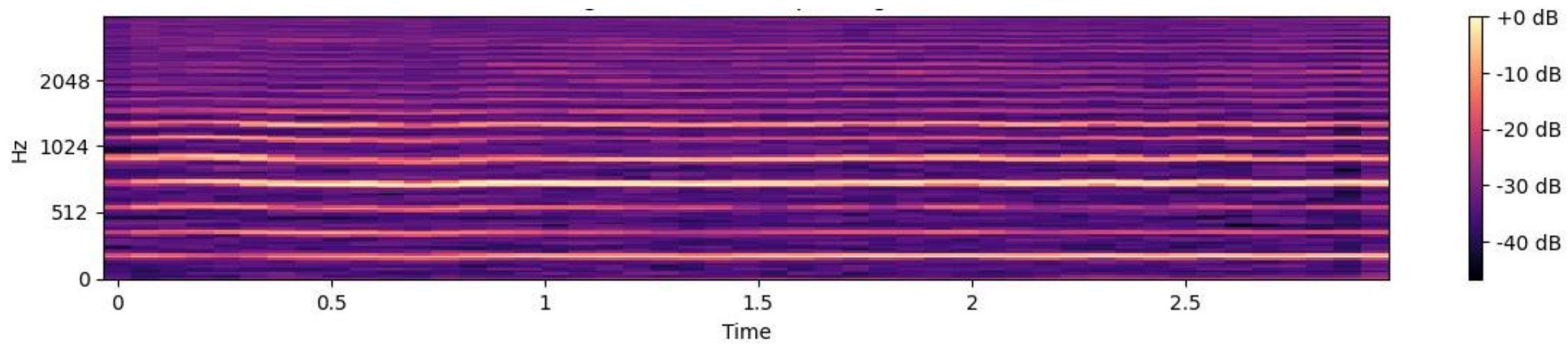

**Figure 3. Log – Mel Spectrogram**

L (t, m) which is a two – dimensional array that represents the Log – transformed Mel spectrogram over the time index t and Mel frequency m, which is the final output of our Log – Mel spectrogram as in figure 3.

After plotting Log – Mel spectrogram Harmonic spectrogram is plotted. An increasing sequence of audible features that infrequently sound unnoticed above a fundamental pitch that is audibly heard is called as a harmonic. A sounding pitch, which often known as the frequency, is the result of our perception of several higher-frequency sound components that are working together to produce the sound we are hearing. This is called as the pitch that we hear is the fundamental. The harmonic spectrum of the sound, which is composed of these higher frequencies or harmonics, which sound above the fundamental, influences the timbre of the sound or tone colour of the sound. Harmonics exist even though they can be challenging to identify as separate elements. In wide – ranging audio sound a harmonic sound is what we understand to be pitched sound, which is what that allows us to recognise the chords and the melodies from the sound.

The audio expression of a sinusoid, or of a horizontal line in a spectrogram representation, is the prototype of a harmonic sound. Alternative way of illustration of what we meant by a harmonic sound is that the sound which is produced by a violin. Once again, the most of the structures seen in the spectrogram are horizontal in character despite being mixed with the components that resemble the noise. The fundamental frequency is exactly multiplied with the whole numbers at harmonic frequencies. The fundamental is the first harmonic, which is followed by the second harmonic, which sounds is at twice the fundamental's frequency, the third harmonic, which sounds is at three times the fundamental's frequency, and so on. So that we can hear the sine waves among the noises. In the audio sample, the voice recordings are same as sine waves that are pure sounds or sound signals, with a harmonic spectrum that contains only one component frequency that of the fundamental. Equation for finding the harmonic spectrogram is given below

Let,

X (k, ω) be Short-Time Fourier transform (STFT) of the audio signal,

H (k, ω) be the harmonic filterbank and,

*H* be the function that extracts harmonic components.

Then,

$$Y(k,\omega) = H(X(k,\omega) * H(k,\omega))$$

------- (11)

Here

Y (k, ω) – Represents the harmonic component spectrogram of audio.

To get harmonic spectrogram of the audio signal get need to get the magnitude of the extracted harmonic components

$$H_{mag}(k,\omega) = | Y(k,\omega) |$$

------- (12)

Logarithmic Transformation is obtained by,

$$H_{log}\,(k,\omega) = log\ log\ \left(1 + \alpha \cdot H_{mag}(k,\omega)\right)$$

------- (13)

α – is the scaling factor.

$H_{mag}$ (k, ω) – Represents the magnitude harmonic component and,

$H_{log}$ (k, ω) – Represents the harmonic spectrogram of the audio signal.

This $H_{log}$ (k, ω) used and plotted against the time index which is then used for our final output of the harmonic spectrogram as in figure 4.

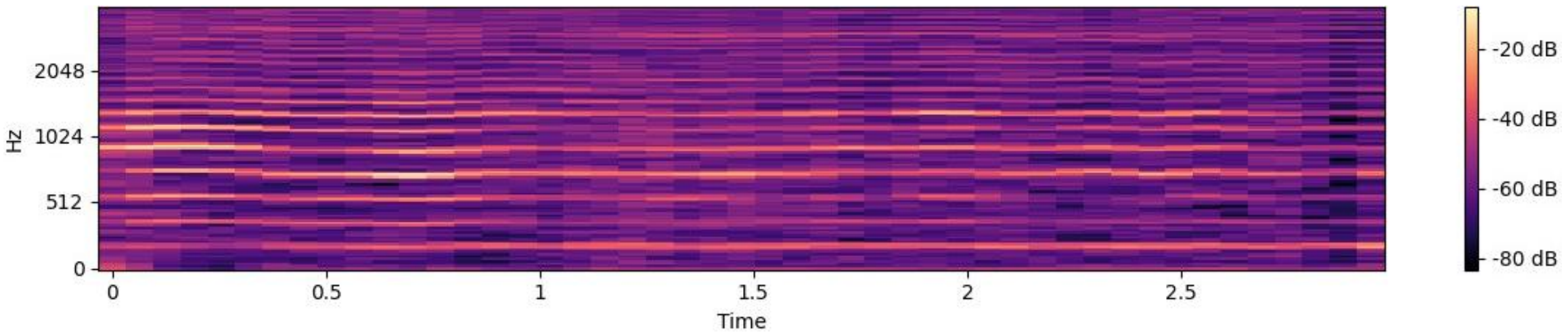

**Figure 4. Harmonic Spectrogram**

After plotting harmonic spectrogram Percussive spectrogram is plotted. The sound which we hear as a click, clap, clash, or knock are all known as the percussive sound. Further common instances include the sound of a drum stroke or a transient that happens during a musical tone's attack phase are also called as the percussive sound.  This percussion is defined as a rhythmic patterning of noise, or a tone attacking noise which is a precise arrangement of pitch and timing, and this is mostly created by membranophone and idiophone musical instruments, as same as by the human body by itself when it blows or inhales.

This percussion audio of a person visualizes the impulse, which is signified by a vertical line in the spectrogram, this is the model of a percussive sound. These percussive components will have the both temporal sparsity and spectral continuity, while harmonic components will have none of these characteristics. Along with the frequency axis, the median filtering is used to build the spectrogram that has been amplified by percussion. The harmonic and percussive time-frequency masks are created by comparing the enhanced spectrograms. This application of harmonic percussive separation can be used to improve the efficiency of audio analysis techniques. The separation is a technique that breaks down a single voice signal into its harmonic and percussion parts by taking the spectrogram image and the vertical and horizontal line into account. The harmonic sound wave have a horizontally smooth time structure, meanwhile the percussion sound wave have a vertically smooth frequency structure in the spectrogram. Equation for percussive spectrogram is same as harmonic spectrogram which is derived below.

Let,

X (k, ω) be short-time Fourier transform (STFT) of the audio signal,

H (k, ω) be the harmonic filterbank and,

P (k, ω) be the percussive filterbank and,

*HPSS* is a Harmonic-Percussive Source Separation function that extracts harmonic and percussive components separately.

Then,

$$P(k,\omega), H(k,\omega) = HPSS\big(X(k,\omega)\big)$$

------- (14)

Here

To get percussive spectrogram of the audio signal get need to get the magnitude of the extracted percussive components

$$P_{mag}(k,\omega) = \mid P(k,\omega) \mid$$

------- (15)

Logarithmic Transformation is obtained by,

$$P_{log}\,(k,\omega) = log\ log\ \left(1 + \alpha \cdot P_{mag}(k,\omega)\right)$$

------- (16)

α – is the scaling factor.

$P_{mag}$ (k, ω) – Represents the magnitude percussive component and,

$P_{log}$ (k, ω) – Represents the percussive spectrogram of the audio signal.

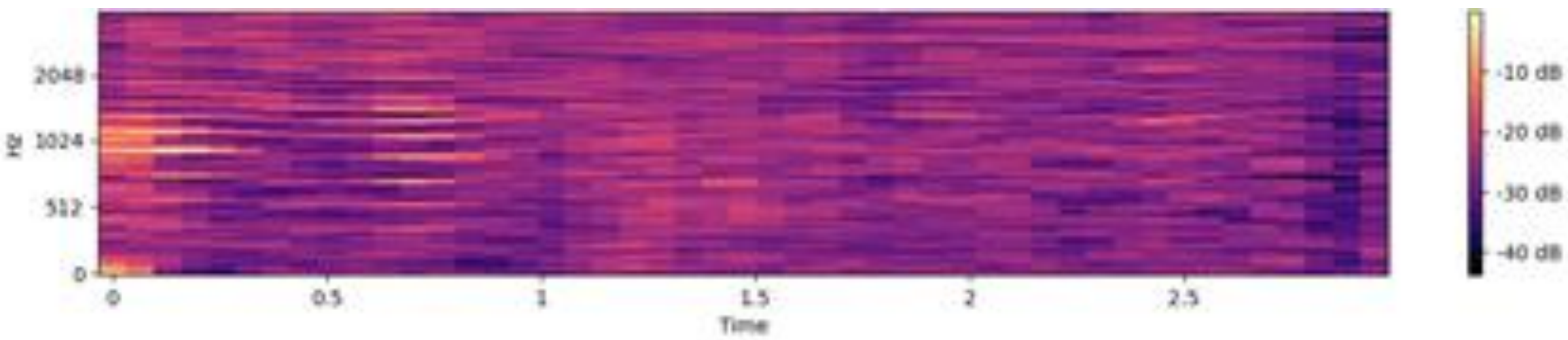

**Figure 5. Percussive Spectrogram**

By these equations percussive spectrogram is plotted for the all-audio samples as in figure 5. All the Log – Mel spectrogram, harmonic spectrogram and percussive spectrogram are plotted for the voice samples and are stored in a separate file. Then these three spectrogram images are analysed to find the difference between the PD and healthy candidates. Then from these three-spectrogram image we plotted Spectral slopes.

**Algorithm 1: Data Pre-Processing**

$Function\ Data_PreProcessing\ (input_folder, output_folder, target_duration = 3.0, target_sr = 256)$

$for\ i \leftarrow 0\ to\ len\ (input_folder)$:

$if\ filename\ ends\ with\ ".wav"\ or\ ".mp3"$:

$audio\ = resample\ (audio, orig_sr = sr, target_sr =$

$$Sample\ Rate\ (SR) = \frac{N_s}{T}$$

$$target_samples\ =\ int\left(target_duration * \frac{sr}{8}\right)$$

$if\ len\ (audio) >\ target\ length:$

$$truncated\ audio = round \left\{\frac{original\ audio}{scaling\ factor}\right\}$$

↲ $end\ if$

$else$

$zeros_to_add\ =\ desireLength\ -\ len(original_samples)$

$original_samples\ +\ [0]\ *\ zeros_to_add$

↲ $end\ else$

$sf.write\ (processed\ audio\ path, audio, sr)$

$Print\ (“processed: \{filename\}”)$

↲ $end\ if$

↲ $end\ for$

↲ $end\ Function$

$Function\ Plot_Log - Mel_Spectrogram\ (processed\ audio\ path):$

$for\ i \leftarrow 0\ to\ len\ (processed\ audio\ path):$

$if\ filename\ ends\ with\ ".wav"\ or\ ".mp3":$

$$M = 2595lo\left(1 + \frac{f}{700}\right)$$

$$M_n[f] = \sum_{f'=0}^{N-1} X(f',t).H_n(f')$$

$$L_n\ [t] = lo\big(M_n(t)\big)$$

$$Y(k,\omega) = H(X(k,\omega) * H(k,\omega))$$

$$H_mag\ (k,\omega) = |\ Y(k,\omega)\ |$$

$$H_log\ \ (k,\omega) = log\ (1 + \alpha \cdot H_mag\ (k,\omega))$$

$$P(k,\omega), H(k,\omega) =\ HPSS(X(k,\omega))$$

$$P_mag\ (k,\omega) = |\ P(k,\omega)\ |$$

$$P_log\ \ (k,\omega) = log\ (1 + \alpha \cdot P_mag\ (k,\omega))$$

$$display.specshow\ (Ln\ [t], Hlog\ (k,\omega), Plog\ (k,\omega))$$

$Print\ (processed: \{filename\})$

↲ $end\ if$

↲ $end\ for$

↲ $end\ Function$

Spectral slopes and energy distribution characteristics are important in spectrogram analysis, especially in the understanding of voice muscle related movement. It is proposed that various spectral patterns could be caused by variations in the kinetics of vocal fold motion. It was shown by a number of examples that the spectral harmonics do not follow earlier conclusions [29]. The formants likewise affect the fundamental's amplitude in the radiated spectrum, but in a predictable manner. Consequently, the influence on the amplitude of the fundamental of the radiated spectrum can be adjusted for given the formant frequencies and the fundamental frequency [30].

So, we plotted spectral slopes and energy distribution graph after analysing all the three spectrogram images. The spectral slopes are plotted separately for the PD audio dataset and plotted separately for the healthy candidate’s audio dataset. Then we analysed these spectral slopes and energy distribution graph and found out that there is a difference between these two spectral slopes graph. The spectral slopes graph of healthy person the amplitude wave is uniformly high and constant initially and it dips towards the end as in figure 7.

While in the spectral slopes graph of PD audio dataset between the two high amplitude waves there exist a low amplitude as you can refer from figure 6. This pattern of low followed by a high amplitude trend was observed in all PD spectral slopes graphs with a varying height of the low amplitude. In some spectral slopes graphs of PD samples only a small amount of low amplitude was able to be observed, whereas in other PD samples we observed a significant difference between the low amplitude and high amplitude.

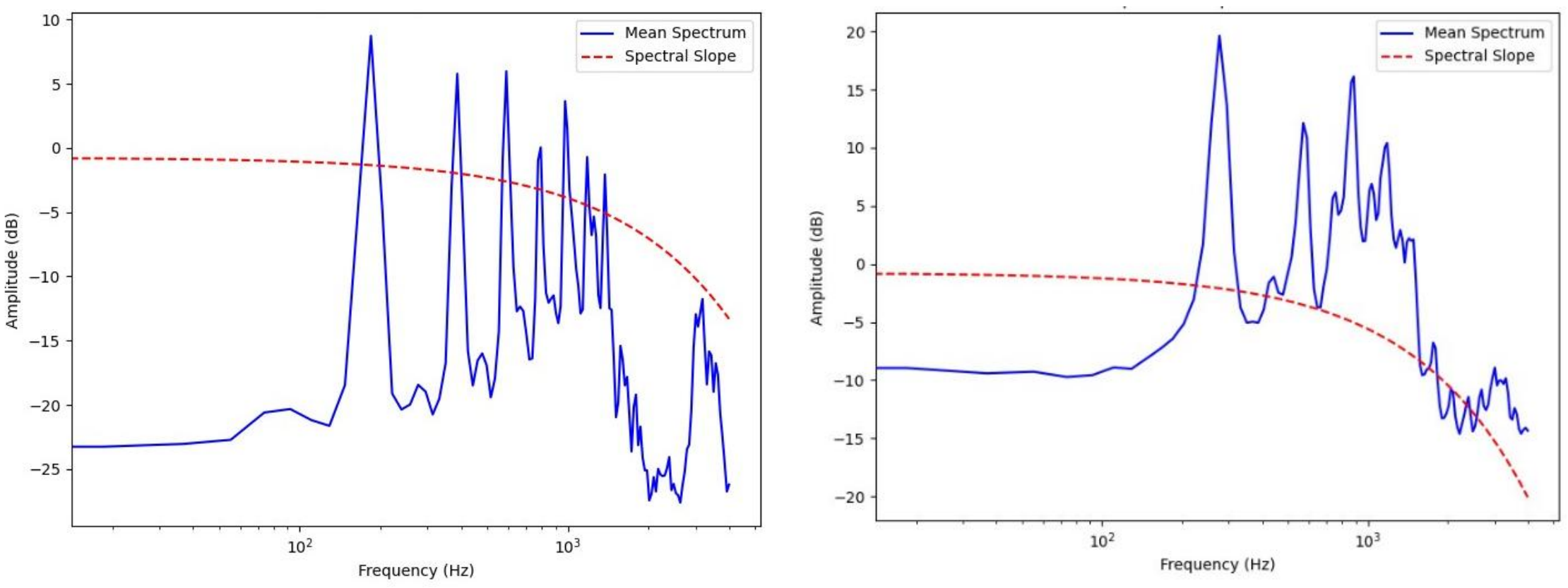

**FIGURE 6. PD spectral slopes graphs**

By referring both figure 6 and 7 these difference in patterns were observed between the PD and HC spectral slopes graphs. Before sending these spectral slopes images into the model we need highlight these important regions of the spectral slope's graphs image. So, we use Gradient-weighted Class Activation Mapping (Grad – CAM) technique which is used to visualize the important region of the spectral slope's graphs image.

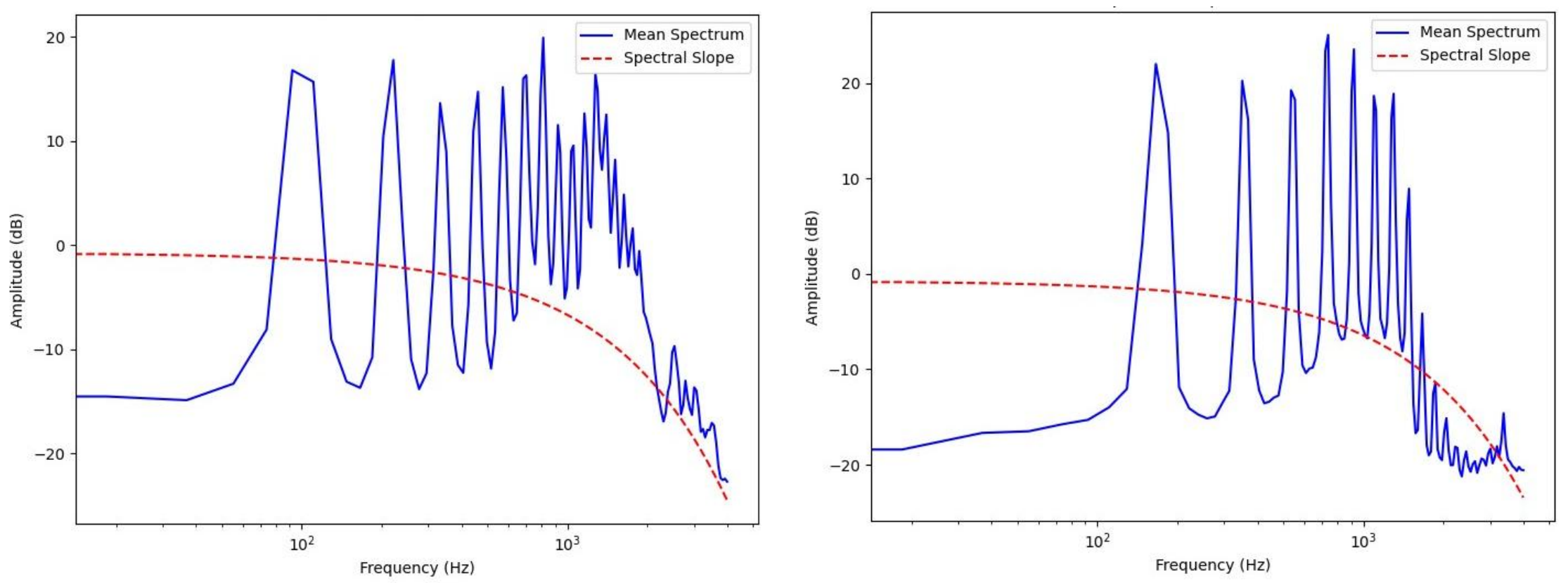

**Figure 7. HC spectral slopes graphs**

### 3.3 Classification

Gradient-weighted Class Activation Mapping (Grad-CAM) is a method used for the localisation of significant regions based on class-specific gradient information. Guided Grad-CAM is a new high-resolution and class-discriminative visualisation that is made possible by combining these localizations with already-existing pixel-space visualisations [31]. By this technique we can highlight the region in which the low amplitudes are found in the PD spectral slopes graphs. It produces a heat map of the spectral slopes graphs and this heat map is then superimposed on the spectral slopes graphs image.

#### *3.3.1 Background of Grad – CAM*

To identify the important part of the image which in turn corresponds to the classification of our model in this we use Grad – CAM technique. We need to input the image as a square image. So, we resize it and as a first step we send the image through the network as a forward pass to find the output class score which is showed in figure 8.

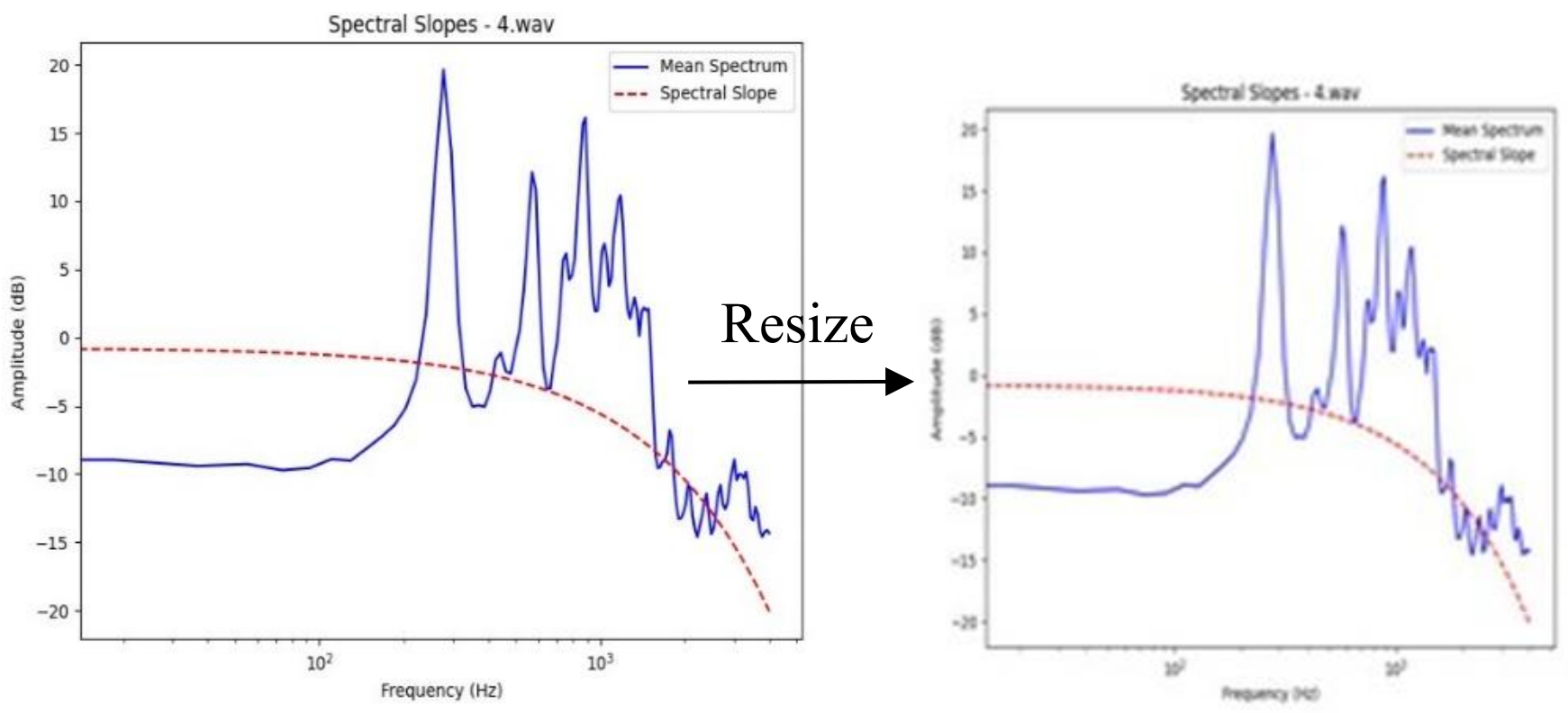

**Figure 8. Resizing the Spectral Slopes**

*Consider a $f()$ which defines forward pass,*

*$X$ represents the input image to the network.*

*$W^k$ represent the $k^{th}$ filter in the last convolutional layer.*

*Then the output of the last convolutional layer is given as,*

$$f(X) = \sum_{k=1}^{K} (W^k * X) + B$$

------- (17)

Here B is the bias term which corresponds to each filter and,

$f(X)$ is final output feature map.

This output $f(X)$ is then given as the input for subsequent layers of the NN. The activation map $A^k$ can be found by just applying the activation function to the output $f(X)$ of the last convolutional layer. The purpose of class activation map visualisation in a CNN is to make that the model is making decisions based on sound reasoning and is free of internal bias resulting from learned spurious correlations or deliberately misleading data selection [38].

Which is given by the equation,

$$A^k = \sigma\left(\sum_{k=1}^{K} (W^k * X) + B\right)$$

------- (18)

$$A^k = \sigma\big(f(X)\big)$$

------- (19)

Here $\sigma$ is the activation function.

By this process of applying the activation function to the output feature map $f(X)$ it is transformed into the activation map $A^k$, in which each element represents the activation level at a particular feature from the input image as captured by the corresponding filter $W^k$.

This activation function $\sigma()$ brings non-linearity into the network model. After finding $A^k$ activation function, it is necessary to calculate the gradient value of the target class score with respect to the feature maps of the last convolutional layer as shown in Grad-CAM architecture from figure 9. So the target class score is calculated by taking the dot product of t feature vector and weights of fully connected layer.

$$S_c = \sum_{i=1}^{N} O_i \,.\, W_{fc_i}$$

------- (20)

Here,

$S_c$ represents target class score,

$O_i$ represents output feature vector,

$W_{fc_i}$ represents weights of the fully connected layer.

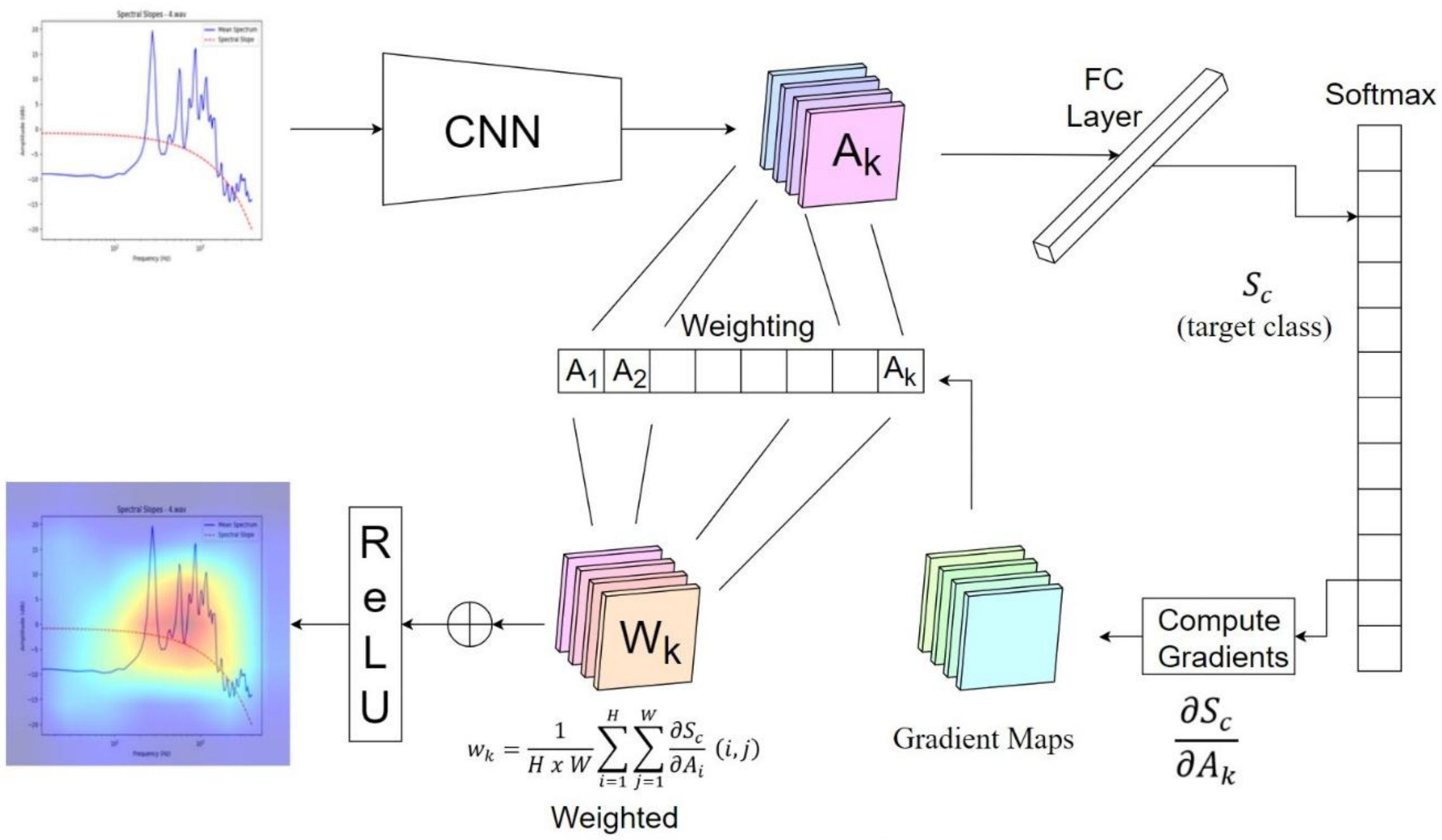

**Figure 9. Grad – CAM Visualization Architecture**

Now we can use the chain rule of calculus to find the gradient of the Target Class Score (TCS). $S_c$ is the activation that maps $A^k$ using equation (21),

$$\frac{\partial S_c}{\partial A_k} = \sum_{i=1}^{N} \frac{\partial S_c}{\partial O_i} \cdot \frac{\partial O_i}{\partial A_i}$$

------- (21)

$\frac{\partial S_c}{\partial O_i}$ represents the grad of the TCS.

$\frac{\partial O_i}{\partial A_i}$ represents the grad of the output feature vector with respect to (w.r.t) the activation feature map (AFM).

$\frac{\partial S_c}{\partial A_k}$ represents grad of the TCS w.r.t AFMs.

As a next step after obtaining the gradient of the TCS w.r.t the AFMs, $\frac{\partial S_c}{\partial A_k}$ , we would need to compute the importance weights as the global average pooling of these gradients.

It is calculated as by the given below formula.

$$w_k = \frac{1}{H\ x\ W} \sum_{i=1}^{H} \sum_{j=1}^{W} \frac{\partial S_c}{\partial A_i}\ (i,j)$$

------- (22)

$H$ represents height of the Activation Map (AM) $A^k$,

$W$ represents the width of AM $A^k$,

$\frac{\partial S_c}{\partial A_k}$ $(i,j)$ represents the gradient at position (i, j) of AM $A^k$,

$w_k$ represents the weights as the global average pooling of these gradients. Finally, we can compute the final Grad-CAM heat map by the use of these importance weights which by linearly combining with the activation maps weighted by their importance. It is found by the equation given below

$$L_c^{Grad-cam} = ReLU\left(\sum_k w_k \, . \, A_k\right)$$

------- (23)

$L_c^{Grad-cam}$ represents the Grad-CAM Heat Map (HM) of input image.

By this step it produces the Grad-CAM HM of input image, which highlights the part of regions from the image that are most applicable for predicting the target class $c$ for the model (refer figure 10).

The Grad-CAM heat map obtained from the input spectral slopes graph further helps the model narrow down on the critical regions that pertain to the predictions made by it. To do so, the areas pertinent to the critical decision-making part are overlaid on the graph image of spectral slopes. The generated heat map-added images are separately saved in a new file and later used for training this model. The customized or fine-tuned Xception model processes these images, generates localization maps through Grad-CAM to indicate regions of interest for concept prediction, and, most importantly, does not apply across all the other CNN model families. Increasing CNN sizes and depth have been pushed further in an attempt to improve the performance of a CNN and are responsible for the increase in computation and storage.

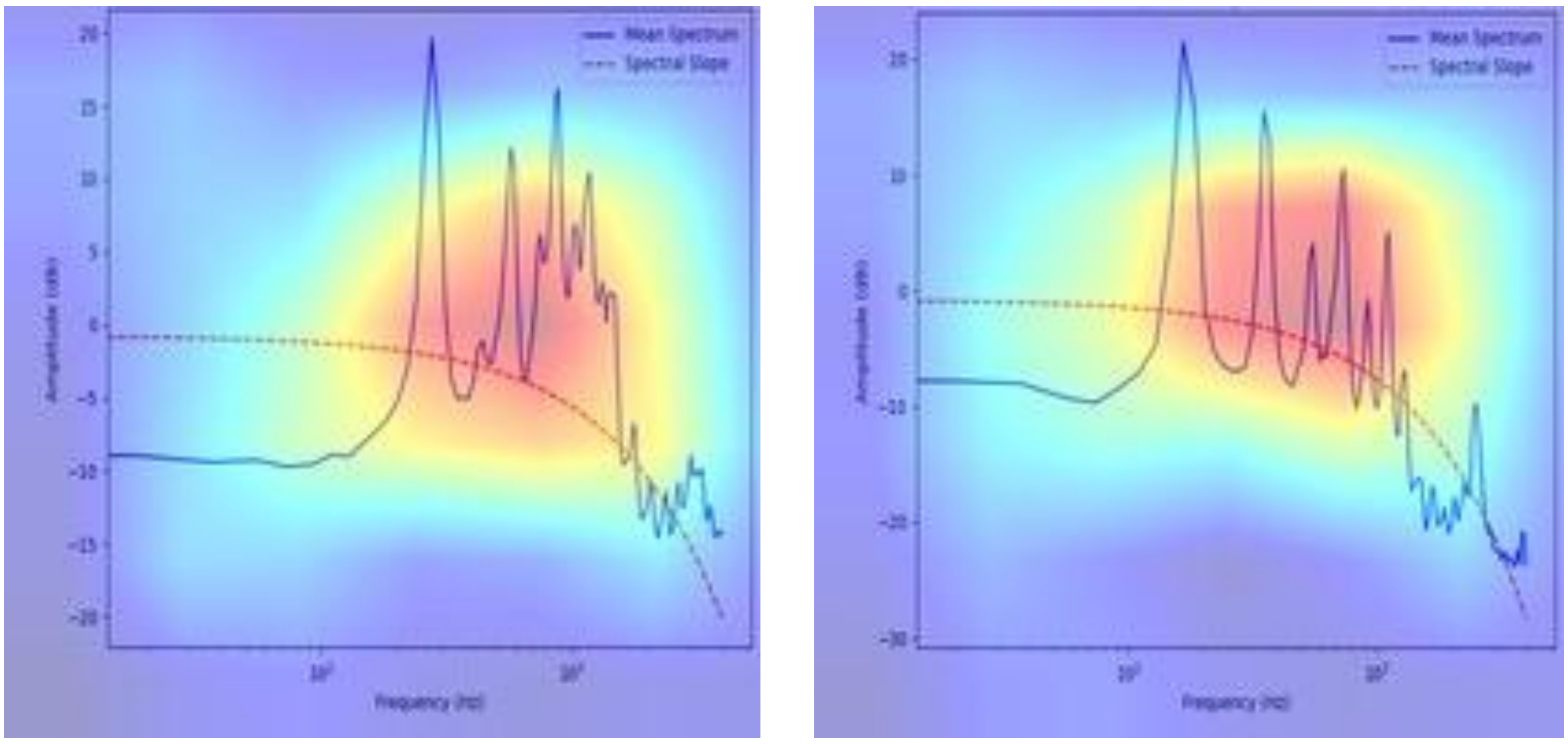

**Figure 10. Grad – CAM Visualised Spectral Slopes**

### *3.3.2 Working of CustNetGC*

CustomNetGC is a CNN architecture using depthwise separable convolutions. The former are more computationally efficient than traditional convolutions. They were first proposed in 2014, and the idea is to separate the filtering and combination stages of convolution. Generalizing this idea, Xception, which is short for "Extreme Inception," first applies filters to individual depth maps and then compresses the input space using a 1×1 convolution. Xception does not apply non-linearity, like in the case of ReLU in some processes unlike Inception. The Xception model typically applies 71 layers to process an image of 299×299 pixels. For the project, however, the images are resized to 244×244 pixels. As a result, some of its parameters have to be adjusted for the needs of the model.

The customizations of the Xception model aim for the more accurate result to achieve reduction in the computational cost. Based on [37], using separable convolutions reduce the size as well as computation overheads for CNNs; there is also a constant try made here for enhancing those in this work too toward Xception. Xception outperforms Inception-v3 by having little accuracy with regards to ImageNet. There are traditional two-step convolution layers: the first is depthwise convolution, where filters the input and the second one is a 1×1 convolution which combines the filtered values to create new features. In this work, fine-tuned parameters of Xception for a project-specific purpose and appropriate input size can help in effective processing with good accuracy.

$$Define\ CustomNet\ architecture$$

$function\ depthwise_separabl_conv(input = (244, 244), filters, kernel_size = (3,3))$:

This CustomNet is a depth – wise separable convolution as same as the Xception model the first step we need to perform is that the input channel is needed to be convolved with a different kernel. This is first step to be performed in a depth wise convolution

$$Depthwise\ convolution$$

$$input = DepthwiseConv2D(kernel_size, padding = 'same')(input) \quad input = BatchNormalization()(input) \quad input = Activation('relu')(input)$$

As a next we perform the point wise convolution that it uses a kernel which iterates through every single pixel of the image by the help of a 1x1 kernel size. In this the depth of the kernel is as same as the number of channel input that the image has.

$$Pointwise\ convolution$$

$$input = Conv2D(filters, (1, 1), padding = 'same')(input) \quad input = BatchNormalization()(input) \quad input = Activation('relu')(input)$$

$$return\ input$$

Then we load our input layer which has a dimension of 244 x 244 and a depth of 4 that is RGBA (Red, Green, Blue and Alpha). The RGBA value is an extension of RGB colour value in which R indicates amount of red present in the image, G indicates amount of green present in the image and B indicates the blue value present in the inputted image. A indicates the opacity of the image. So to train out model we don't need this Alpha value as it remains same for all. So, we are removing this value from the image data to reduce the model's complexity. To remove the Alpha value we use the below formula

$$new_Channel_Code = C + ((255 - C) * A)/255$$

------- (24)

Where,

C is the existing Channel code,

A is the Alpha value of the particular pixel.

$$Input\ layer$$

$$input_layer = Input(shape = input_shape(244,244))$$

$$new_R = R + ((255 - R) * A)/255$$

$$new_G = G + ((255 - G) * A)/255$$

$$new_B = B + ((255 - B) * A)/255$$

So by this formula we removed the alpha value from the image and the image is loaded to the model which is been trained. Then we add the entry flow of our model.

$$Entry\ flow$$

$$x = Conv2D(32,(3,3), strides = (2,2), padding = 'same')(input_layer)x = BatchNormalization()(x)x = Activation('relu')(x)x = Conv2D(64,(3,3), padding = 'same')(x)x = BatchNormalization()(x)x = Activation('relu')(x)$$

This entry flow in the network model helps to extract features from the inputted image by a series of convolutional and pooling operations. This entry flow step gradually increases the level of construction of the model. Then it undergoes the middle flow.

$$Middle\ flow$$

$$for\ _\ in\ range(num_middle_blocks):$$

$$residual = x$$

$$x = depthwise_separable_conv(x, 128, (3,3)) \quad x = depthwise_separable_conv(x, 128, (3,3)) \quad x = depthwise_separable_conv(x, 128, (3,3)) \quad x = Add()([residual, x])$$

The Middle flow is just a series of repeated blocks of code in for loop, each performs a depth wise separable convolution. This step is done to help the model to capture a higher – level of features and to make the model to learn more of the complexity of the inputted image data. Then it finally it goes through the exit flow.

$$Exit\ flow$$

$$residual = Conv2D(1024,(1,1), strides = (2,2), padding = 'same')(x)residual = BatchNormalization()(residual)x = depthwise_separable_conv(x, 244, (3,3))x = depthwise_separable_conv(x, 244, (3,3))x = MaxPooling2D((3,3), strides = (2,2), padding = 'same')(x)x = Add()([residual, x]$$

$$x = depthwise_separable_conv(x, 728, (3,3))x = depthwise_separable_conv(x, 728, (3,3))x = MaxPooling2D((3,3), strides = (2,2), padding = 'same')(x)x = Add()([residual, x])$$

In the exit flow batch normalization is done and followed by global average pooling or max pooling 2D is done to refine the extracted feature and to make them prepared for the final prediction. By these the output layer is formed for the model which is a dense layer.

$$Output\ layer$$

$$output = GlobalAveragePooling2D()(x)output = Dense(num_classes, activation = 'softmax')(output)$$

Finally, the model is created by the customized input image size and parameters. Which is used for making prediction for our work.

$$Create\ model$$

$$model\ =\ Model(inputs = input_layer, outputs = output)$$

After creation of the model the model is been trained and it is evaluated against the evaluation metrics.

**Algorithm 2: Grad – CAM**

$\boldsymbol{Function}\ grad_cam(model, img_arr, layerName, index, size = (244, 244))$

$img\ =\ preprocessing.image(imagePath, target_size = (244, 244))$

$img\ =\ tf.keras.preprocessing.image.img_to_array(img)$

$img\ =\ np.expand_dims(img, axis = 0)$

$X = img$

$$f(X) = \sum_{k=1}^{K} (W^k * X) + B$$

$$A^k = \sigma\left(\sum_{k=1}^{K} (W^k * X) + B\right)$$

$$S_c = \sum_{i=1}^{N} O_i \, . W_{fc_i}$$

$$\frac{\partial S_c}{\partial A_k} = \sum_{i=1}^{N} \frac{\partial S_c}{\partial O_i} \cdot \frac{\partial O_i}{\partial A_i}$$

$$w_k = \frac{1}{H\ x\ W} \sum_{i=1}^{H} \sum_{j=1}^{W} \frac{\partial S_c}{\partial A_i}\ (i, j)$$

$$L_c^{Grad-cam} = ReLU\left(\sum_k w_k\ .\ A_k\right)$$

$heatmap\ = L_c^{Grad-cam}$

$\boldsymbol{return} \leftarrow\ heatmap$

↲ $\boldsymbol{end\ Function}$

**Algorithm 3: CustomNet**

```
function depthwise_separable_conv(input, filters, kernel_size):
    input = DepthwiseConv2D(kernel_size, padding = 'same')(input)
    input = BatchNormalization()(input)
    input = Activation('relu')(input)

    input = Conv2D(filters, (1, 1), padding = ' same')(input)
    input = BatchNormalization( )(input)
    input = Activation('relu')(input)
return ← input
    input_layer = Input(shape = input_shape(256,256))
    new_R = R + ((255 − R) ∗ A)/255
    new_G = G + ((255 − G) ∗ A)/255
    new_B = B + ((255 − B) ∗ A)/255
    x = Conv2D(32, (3, 3), strides = (2, 2), padding = 'same')(input_layer)
    x = BatchNormalization()(x)
    x = Activation('relu')(x)
    x = Conv2D(64, (3, 3), padding = 'same')(x)
    x = BatchNormalization()(x)
    x = Activation('relu')(x)
    for i ← in range(num_middle_blocks):
        residual ⇐ x
          x = depthwise_separable_conv(x, 128, (3, 3))
          x = depthwise_separable_conv(x, 128, (3, 3))
          x = depthwise_separable_conv(x, 128, (3, 3))
          x = Add()([residual, x])
    ↲ end for
    residual ⇐ Conv2D(1024, (1, 1), strides = (2, 2), padding = 'same')(x)
    residual ⇐ BatchNormalization()(residual)

    x = depthwise_separable_conv(x, 256, (3, 3))
    x = depthwise_separable_conv(x, 256, (3, 3))
    x = MaxPooling2D((3, 3), strides = (2, 2), padding = 'same')(x)
    x = Add()([residual, x])
```

**Integrated with CatBooost**

$$for\ i \leftarrow 1\ to\ len\ (number\ of\ nodes):$$
$$for\ j \leftarrow 1\ to\ i:$$
$$M\ \leftarrow\ LearnOneTree\ (Xj, Yj)$$
$$\text{↲}\ end\ for$$
$$Mi\ \leftarrow\ Mi\ +\ M$$
$$\text{↲}\ end\ for$$
$$Return \leftarrow\ M1 \ldots Mn;$$

## 4.PERFORMANCE EVALUATION

### 4.1 Implementation Platform of proposed Model

The proposed model used for voice based Parkinson disease detection is implemented by using python programming. Tensorflow, keras, numpy etc. Are the libraries used for implementing our work.

### 4.2 Evaluation Metrics

After pre-processing the audio dataset, the L-mHP feature-a combination of Log-Mel spectrogram, harmonic spectrogram, and percussive spectrogram-was computed. Analysis of these spectrograms and their patterns showed a clear spectral slope characteristic in Parkinson's Disease (PD) voice samples: a low-amplitude wave repeating between high-amplitude waves. For the model to focus on this prominent pattern, an attention mechanism was used through Grad-CAM. It identified significant regions in the spectral slopes graph through a generated heat map in which red color is shown to have high importance decision regions and less important regions being marked by the blue color. The heat maps were overlaid on the spectral slope graphs of PD and normal voice samples, respectively. Next, the overlaid spectral slope images were split into training and validation datasets.

To upload these images to the model, a fine-tuned custom model, here named CustomNet, was developed, taking in input images of size 244×244 with specific stride values. The Grad-CAM visualizations from both classes were fed into this model. Using categorical boosting technique called CatBoost further enhanced the model named CustNetGC Boosted. Once well trained, testing the model was carried out using standard metrics. While there are plenty of performance metrics available in the deep learning community, selecting the most appropriate one to evaluate classifier performance is difficult [41]. One of the key evaluation metrics used here is accuracy, which was calculated as correct predictions over the total predictions and multiplied by 100 to be expressed as a percentage. Accuracy is calculated by the following formula,

$$Number_of_Correct_Predictions = TP + FN$$

$$Total_Number_of_Predictions = TP + TN + FP + FN$$

From above two equation we get,

$$Accuracy = \frac{TP + FN}{TP + TN + FP + FN} \times 100\%$$

------- (25)

Here,

$TP$ represents True Positives, $TN$ represents True Negatives, $FP$ represents False Positives and $FN$ represents False Negatives present in the model.

Then confusion matrix is plotted to showcase the presentation of our classification model on the given set of validation data for which the true values are found. The confusion matrix is a 2 X 2 matrix using the components of TP, FP, TN, and FN. From the confusion matrix, we can get to know about various performance metrics can be derived with these metrics we can provide insights of the model's performance from a different perspective.

Precision of the model is evaluated using the formula is given below as

$$Precision = \frac{TP}{TP + FP}$$

------- (26)

Similarly specificity of the model is evaluated using the formula is given below as

$$Specificity = \frac{TN}{TN + FP}$$

------- (27)

For recall is given as

$$Recall = \frac{TP}{TP + FN}$$

------- (28)

Next to get F1 score of the model, we use the formula given below as

$$F1\, score = 2\, X\, \frac{Precision\;\; X\, Recall}{Precision\;\; + Recall}$$

------- (29)

As a next evaluation metrics Receiver Operating Characteristic (ROC) curve is plotted. This curve illustrates the diagnostic ability of a binary classification model across various threshold values. This ROC curve is plotted as keeping the X – axis as the False Positive Rate or it can be denoted as 1 – Specificity and the Y – axis as True Positive Rate or also known as Sensitivity. It is computed as

$$ROC = \left(\frac{FP}{FP + TN}, \frac{TP}{TP + FN}\right)$$

------- (30)

The Area Under the ROC Curve formula is given by,

$$AUC = \left(\frac{(b1 + b2)Xh}{2}\right)$$

------- (31)

Then False Positive Rate (FPR) is a performance metric which used to find the number of false data which are predicted as true for the given data set can found. This is a complementary of specificity. Its equation is given below.

$$FPR = \frac{FP}{FP + TN}$$

------- (32)

The Thresholds metrics of the model is used to identify the instances into positive or negative classes by the basis of the predicted probabilities output by a model. By this we can be able to identify the point at which predicted probabilities are found to be considered as positive predictions.

**Algorithm 4: Evaluation metrics**

$$Function\ evaultion_metrics(TP, TN, FP, FN)$$

$$Accuracy = \frac{Total_Number_of_Predictions}{Number_of_Correct_Predictions} \times 100\%$$

$$Accuracy = \frac{TP+TN}{TP+TN+FP+FN} \times 100\%$$

$$Specificity = \frac{True\ Negatives}{True\ Negatives+False\ Positives}$$

$$Specificity = \frac{TN}{TN+FP}$$

$$for\ i \leftarrow 0\ to\ 1:$$

$for\ j \leftarrow 0\ to\ 1:$

$Confusion[0][0] = TP$

$Confusion[0][1] = FP$

$Confusion[1][0] = TN$

$Confusion[1][1] = FN$

↲ $end\ for$

$$Sensitivity = \frac{True\ Positives}{True\ Positives+False\ Negatives}$$

$$Sensitivity = \frac{TP}{TP+FN}$$

$$F1\ score\ = 2\ X\ \frac{Precision\ \ X\ Recall}{Precision\ \ +Recall}$$

$$Precision\ = \frac{True\ Positives}{True\ Positives+False\ Positives}$$

$$Precision\ = \frac{TP}{TP+FP}$$

$$FPR\ = \frac{False\ Positives}{False\ Positives+True\ Negatives}$$

$$FPR\ = \frac{FP}{FP+TN}$$

$$ROC\ = \left(\frac{False\ Positives}{False\ Positives+True\ Negatives}, \frac{True\ Positives}{True\ Positives+False\ Negatives}\right)$$

$$ROC\ = \left(\frac{FP}{FP+TN}, \frac{TP}{TP+FN}\right)$$

$$AUC = \left(\frac{(b1+b2)Xh}{2}\right)$$

$$\downarrow end\ Function$$
$$return \leftarrow metrics$$

## 5.RESULT AND DISCUSSIONS

### 5.1 Results Analysis

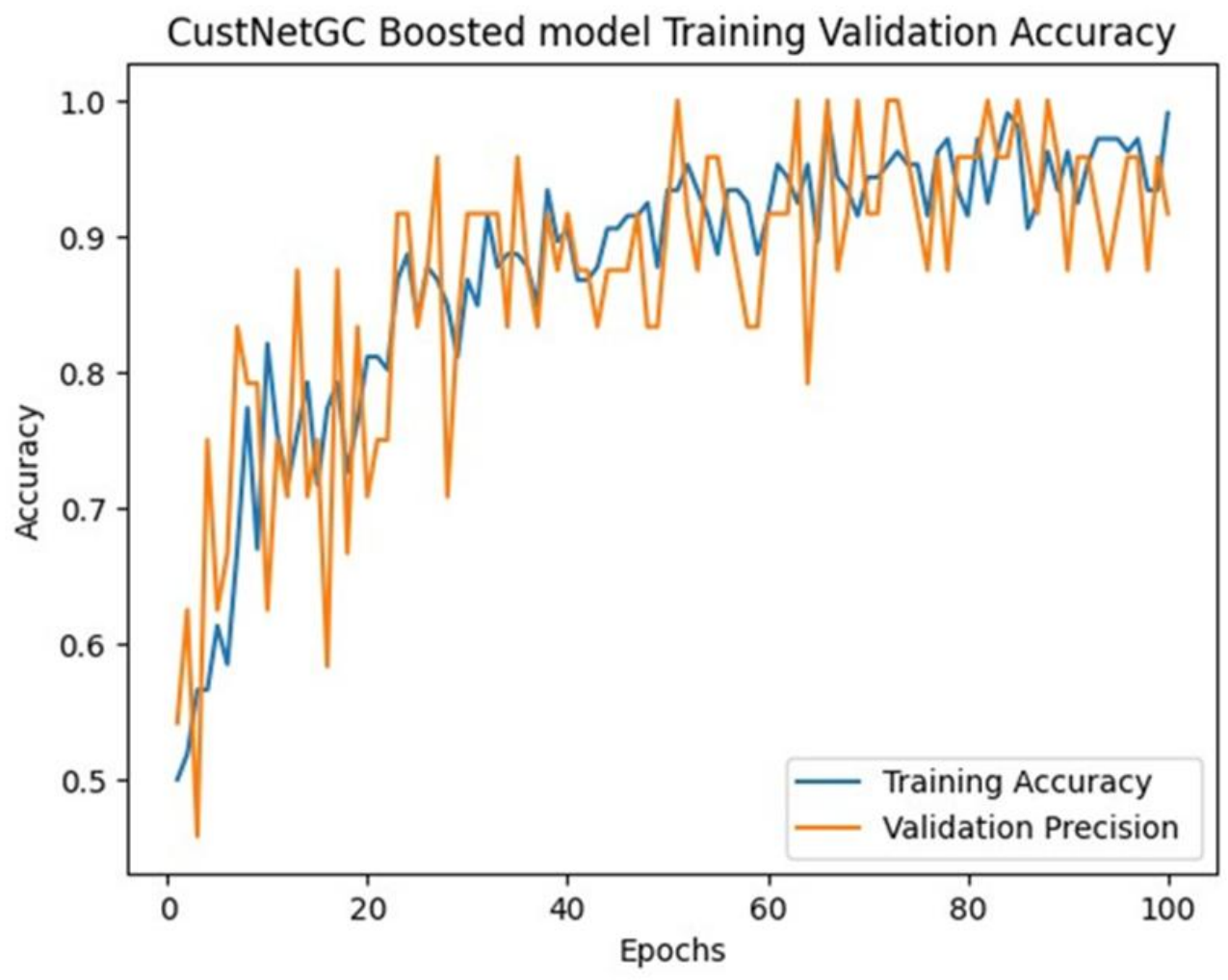

**Figure 11. Boosted Model Accuracy**

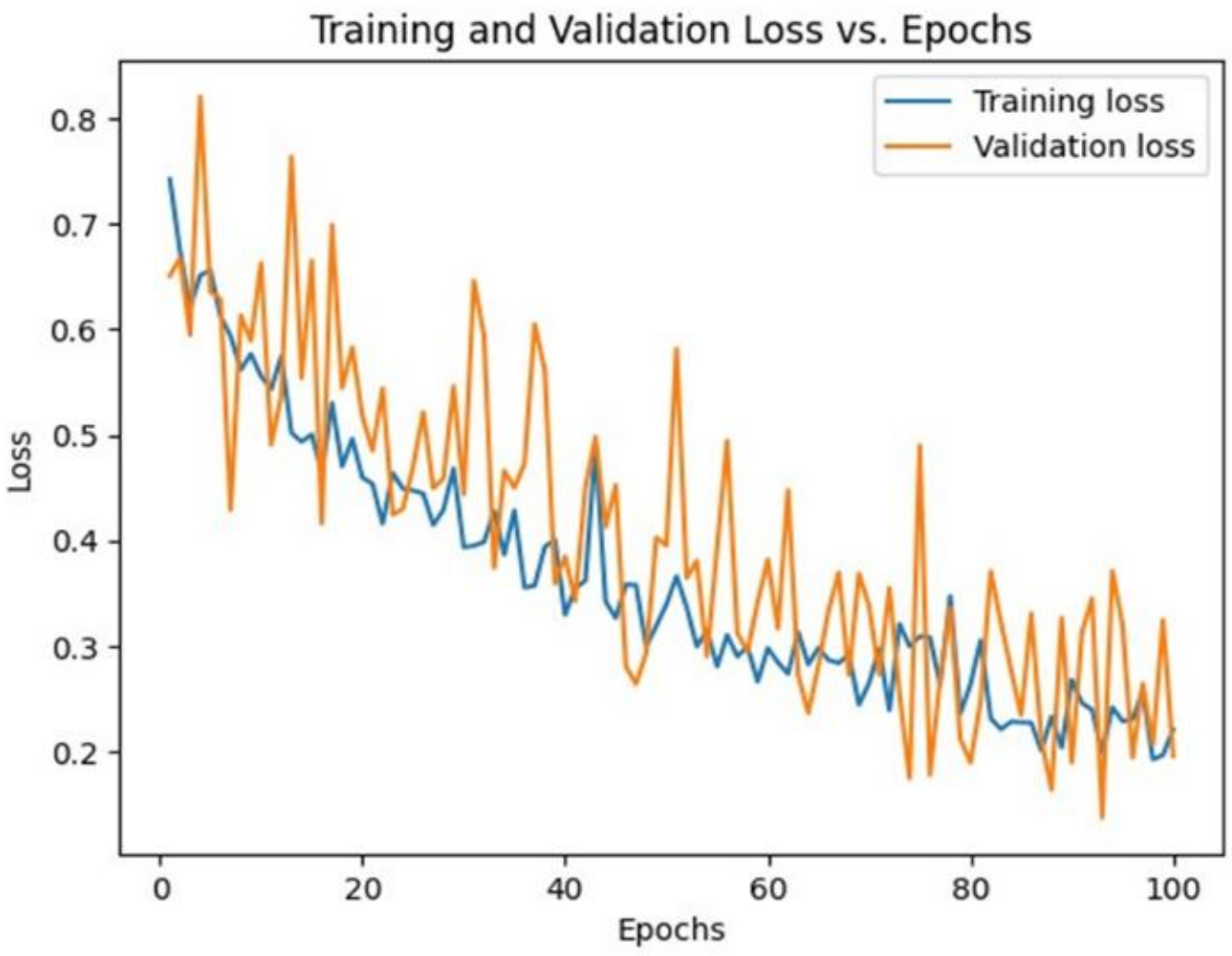

**Figure 12. Boosted Model Loss**

The CustNetGC Boosted model had achieved the training accuracy at 0.9906, which translates into excellent performance for most of the training examples. For evaluating the model using the unseen data, a confusion matrix was plotted using a validation dataset containing 162 samples with equal samples being taken from each of the PD class and the remaining in the HC class. The model correctly classified 80 PD samples as PD-affected voices and 78 HC samples as healthy voices. In this, it incorrectly classified 4 HC samples as PD-affected voices.

The F1-Score of the model was 0.8459, indicating that it has found a good trade-off between precision and recall. It also reflected precision to be at 0.9583 with high accuracy for identification of PD samples. Using these measures, the model's specificity and recall were computed. In addition, the Precision-Threshold curve and the ROC curve were graphed. The AUC for the PD class was 0.90 and that of the HC class was 0.89, thus verifying the model's excellent performance on the task of discriminating between the two classes.

Training loss is the error during training; validation loss corresponds to the unseen data from the validation set about the performance of the model. The training procedure for the CustNetGC model showed a training loss that dropped with an increasing number of epochs, which proves that the model is learning effectively as the number of epochs increases. Similarly, the validation loss also shows a downward trend, indicating generalization of the model to the unseen data without overfitting or underfitting.

The CustNetGC Boosted model indicates stable training with consistent decline in both training and validation losses across epochs, thereby achieving reliable performance on both known and unseen datasets.

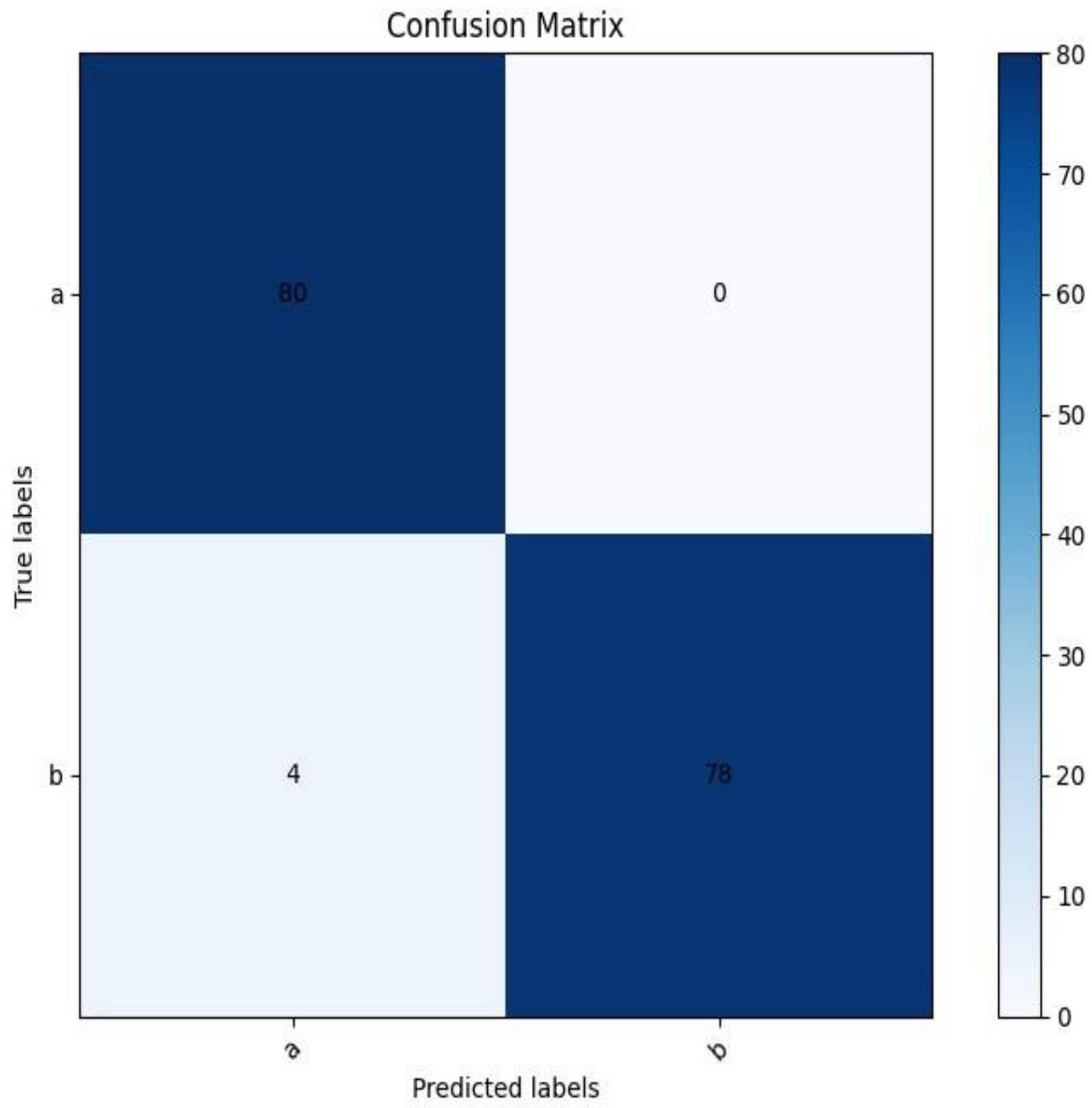

**Figure 13. Confusion Matrix**

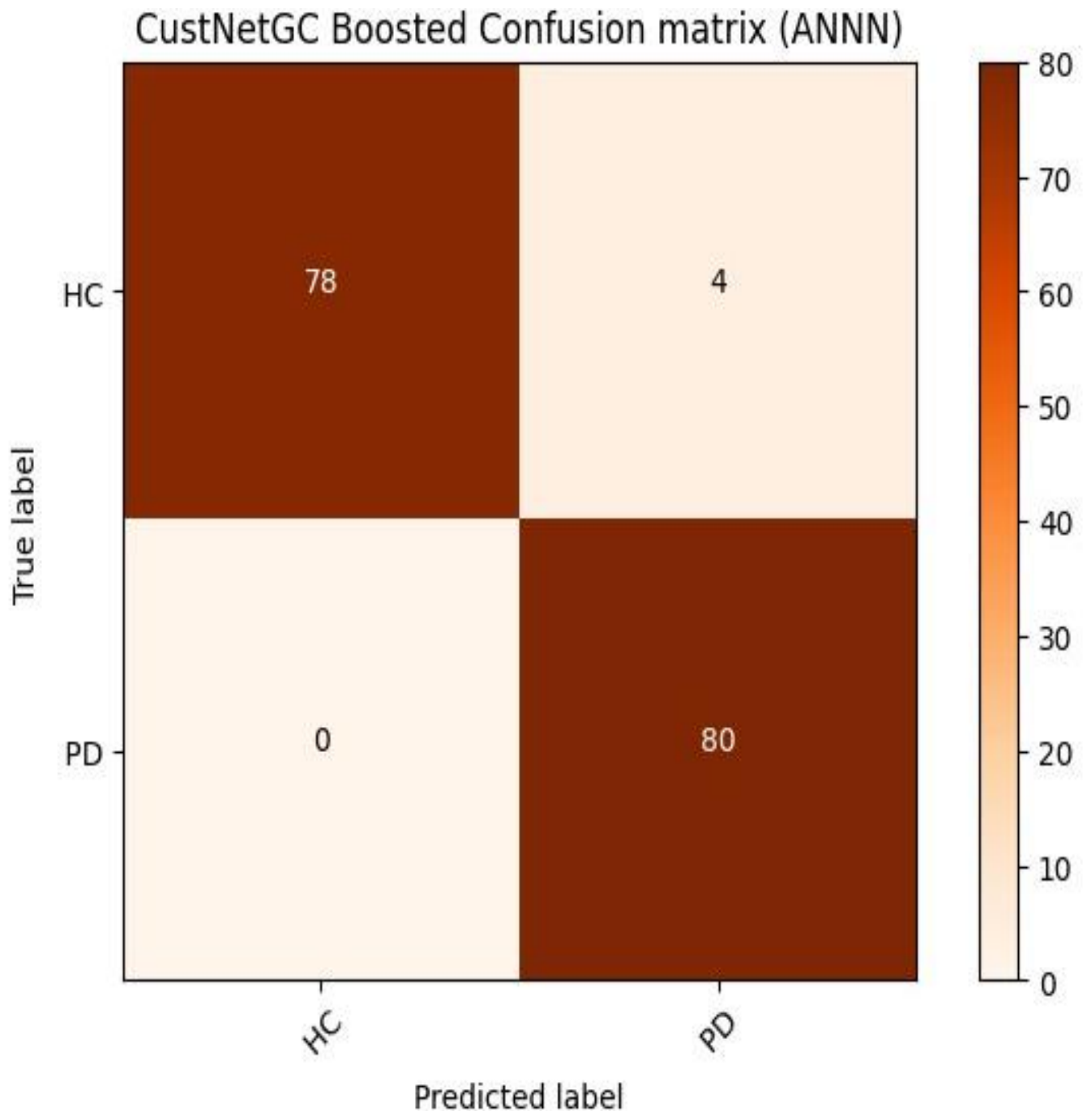

**Figure 14. Confusion Matrix(ANNN)**

A confusion matrix is a tool used to evaluate the performance of our CustNetGC model on a separate validation dataset, where the true values are known. It provides a detailed visualization of how well the model predicts each class.

- Columns in the matrix represent the predicted instances for each class.
- Rows in the matrix represent the true instances for each class.

This matrix highlights the correctness and errors in the model's predictions, offering valuable insights into its performance.

In addition to the confusion matrix, other performance metrics such as F1 score and precision are used to quantify the model's ability to correctly classify PD and HC classes. These metrics provide a comprehensive evaluation of the CustNetGC model's performance.

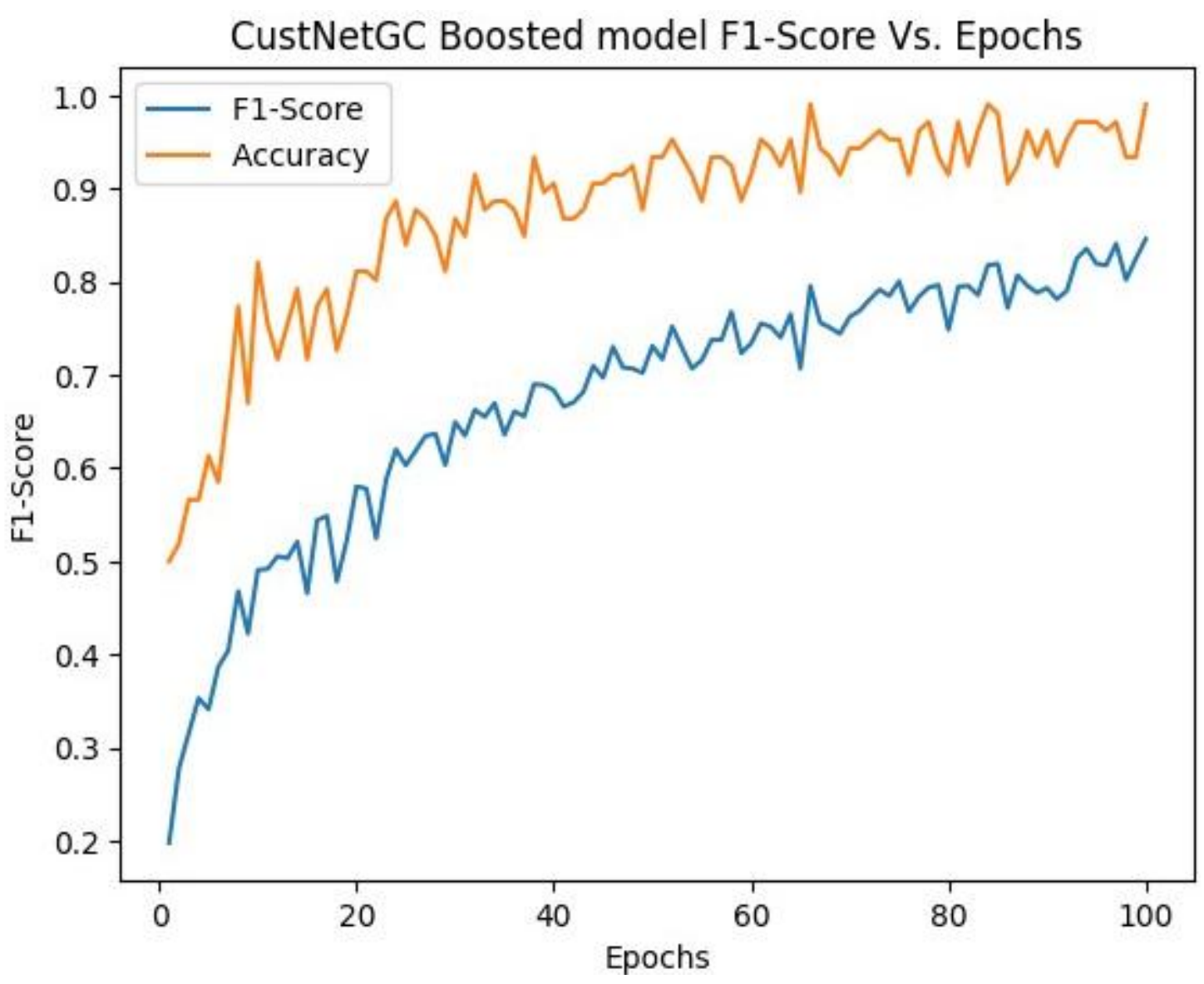

**Figure 15. Boosted model F1 - Score**

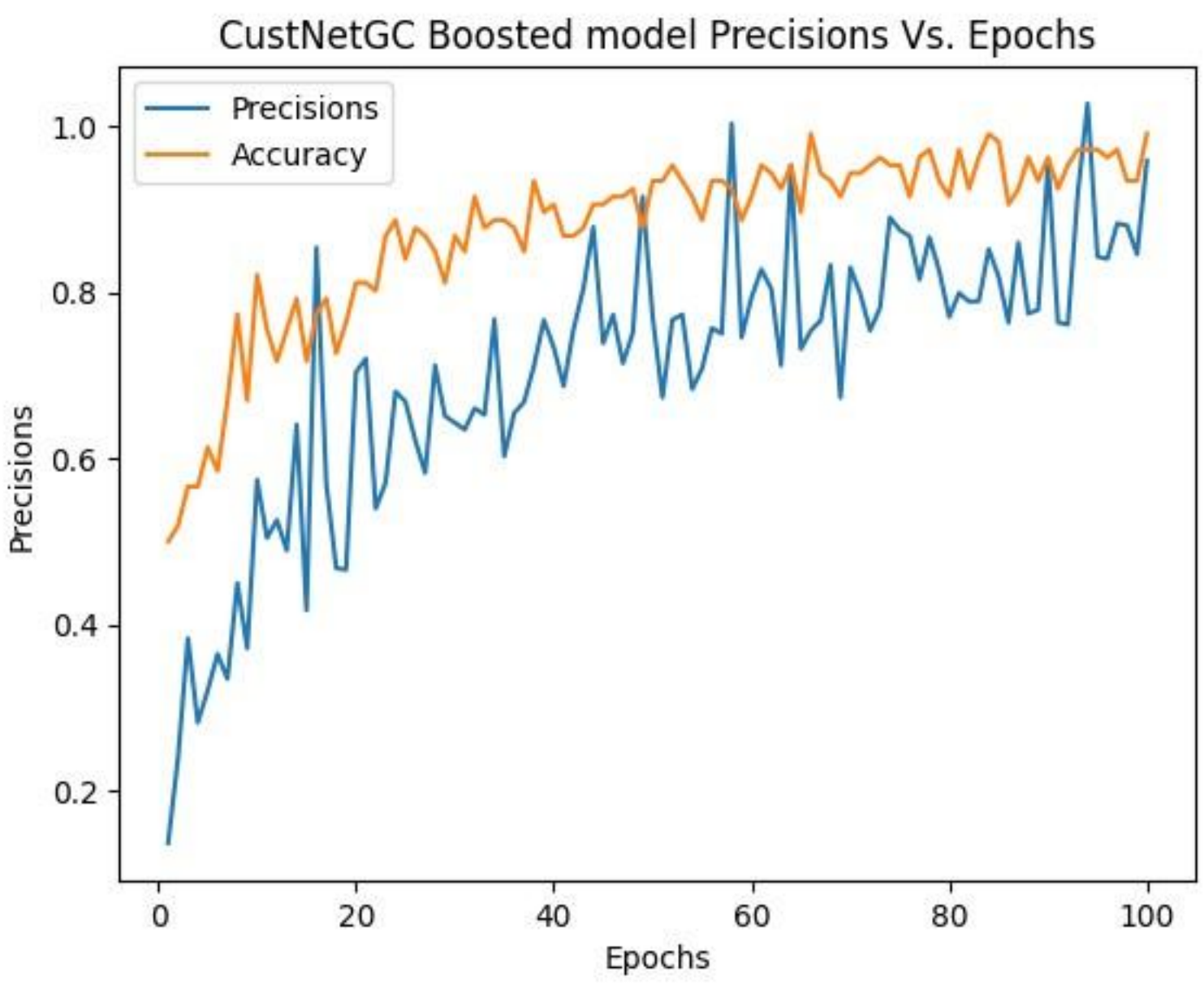

**Figure 16. Boosted model Precision**

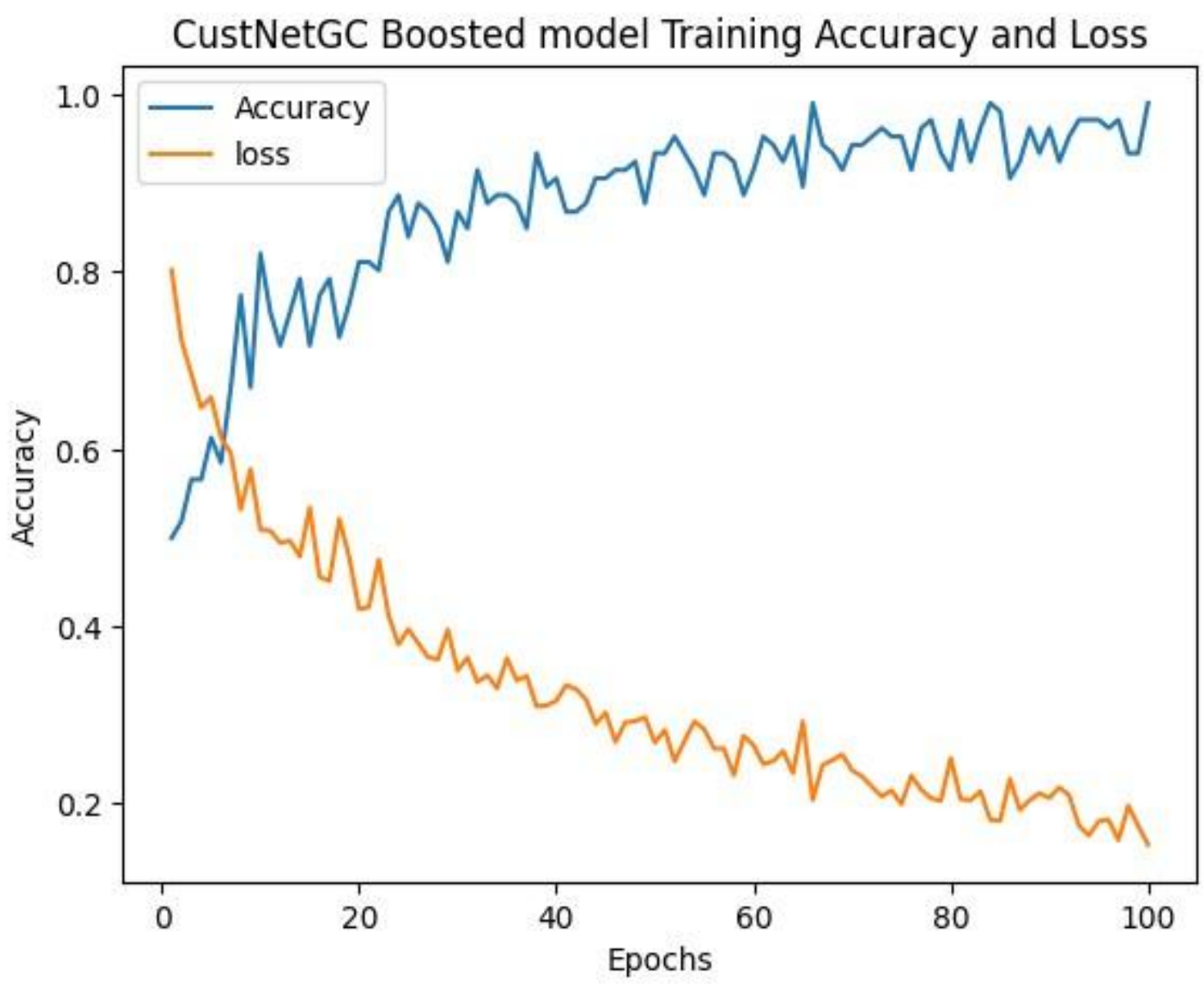

**Figure 17. Training Accuracy and Loss**

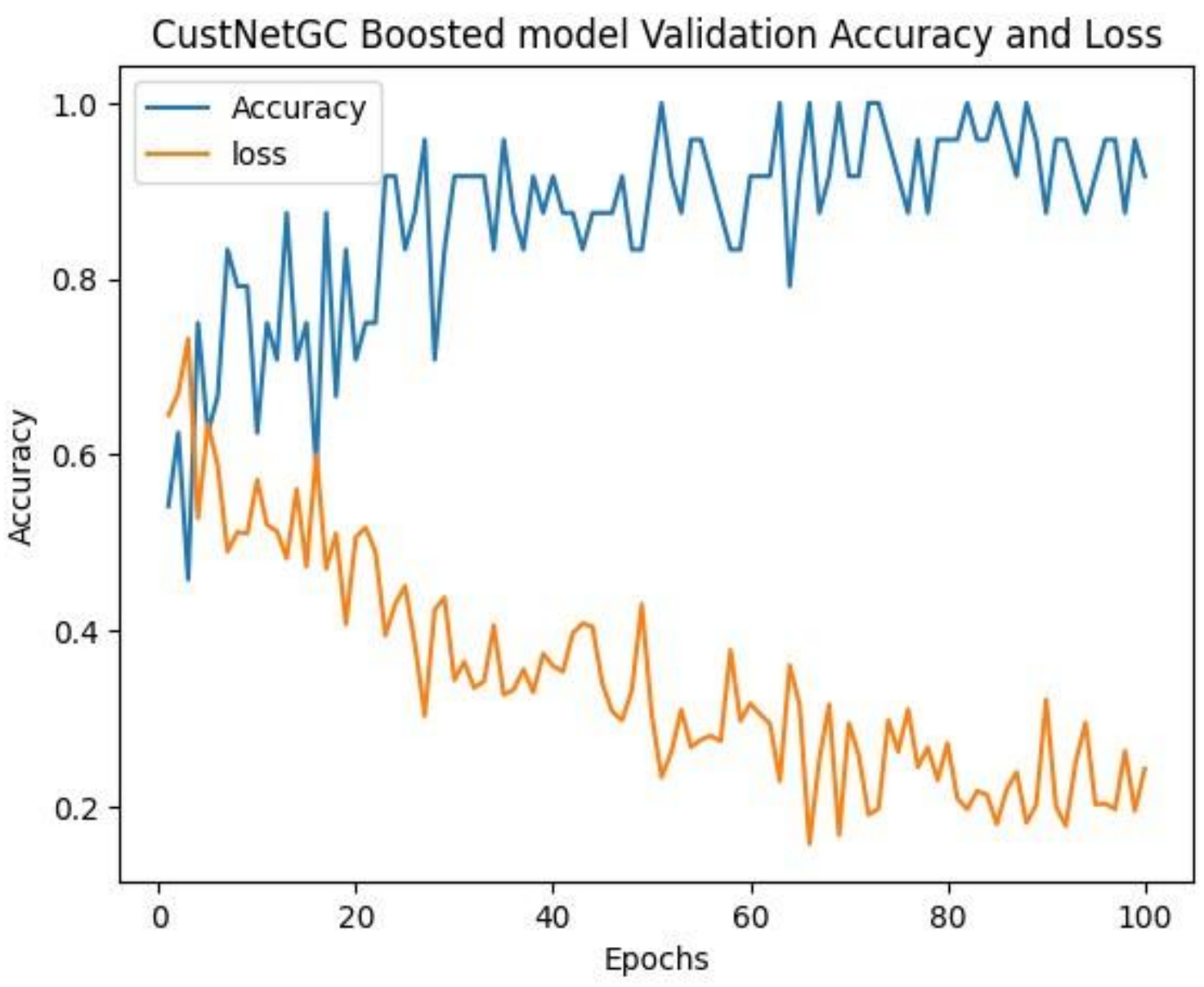

**Figure 18. Validation Accuracy and Loss**

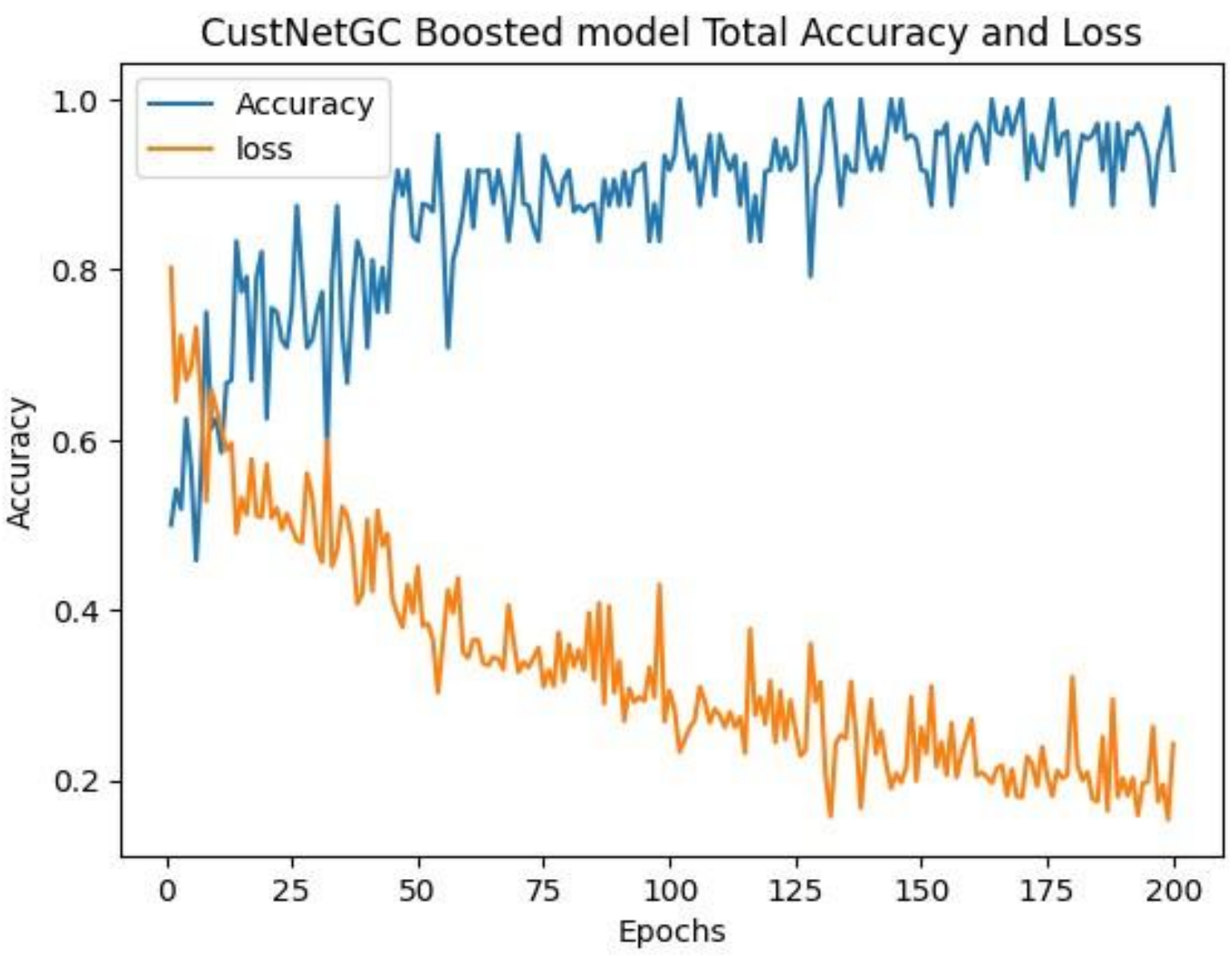

**Figure 19. Total Accuracy and Loss**

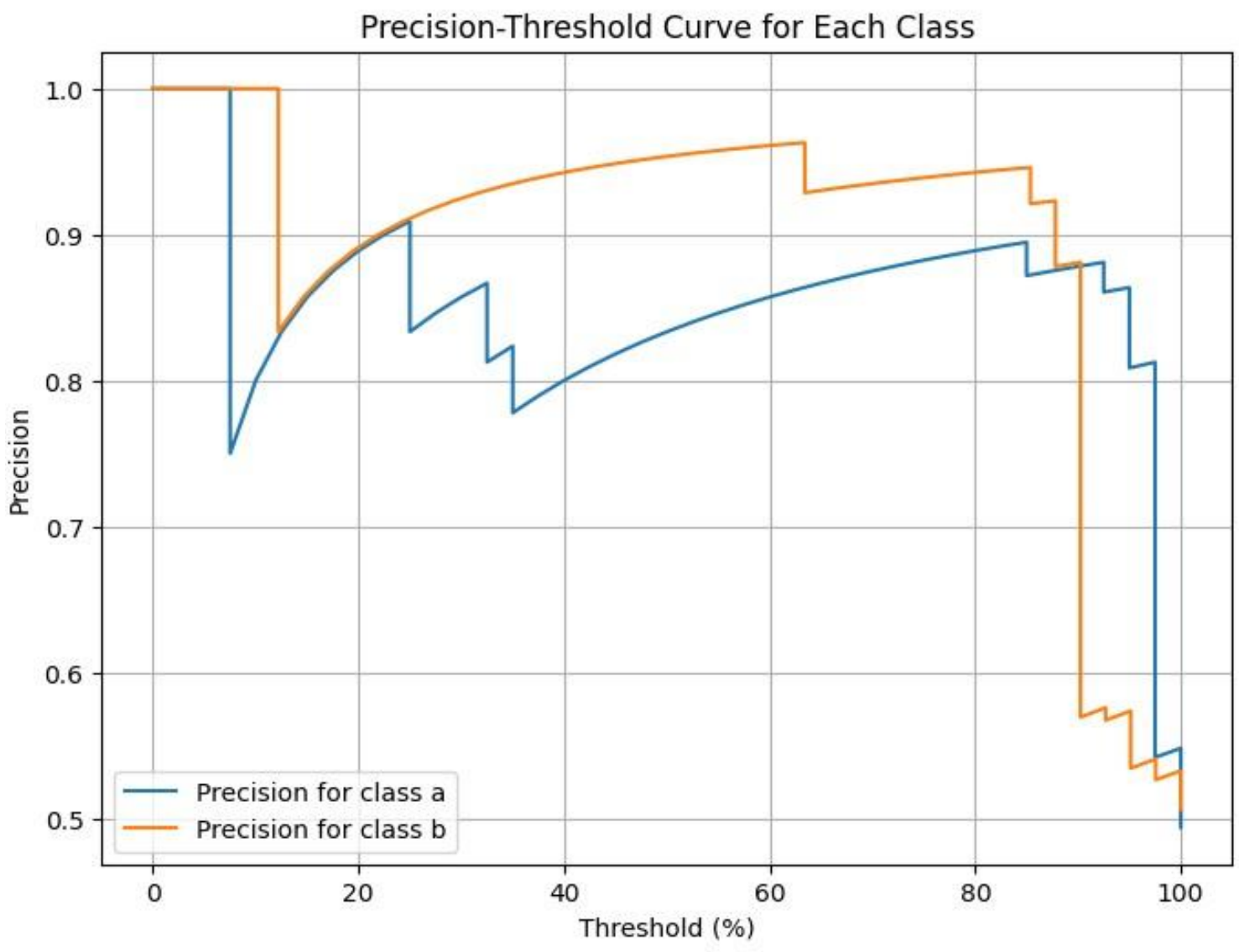

**Figure 20. Precision vs. Threshold Curve**

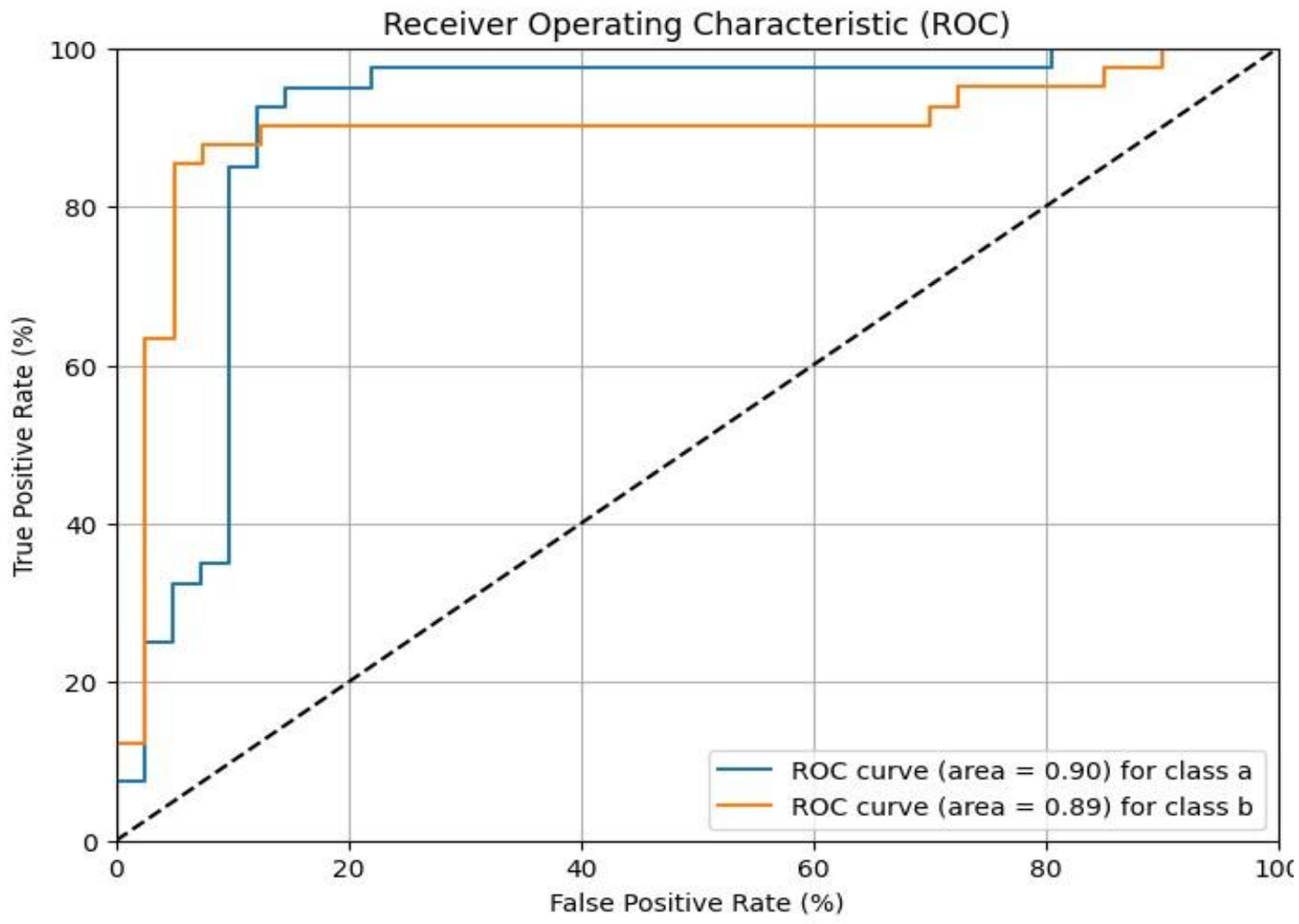

**Figure 21. ROCAUC**

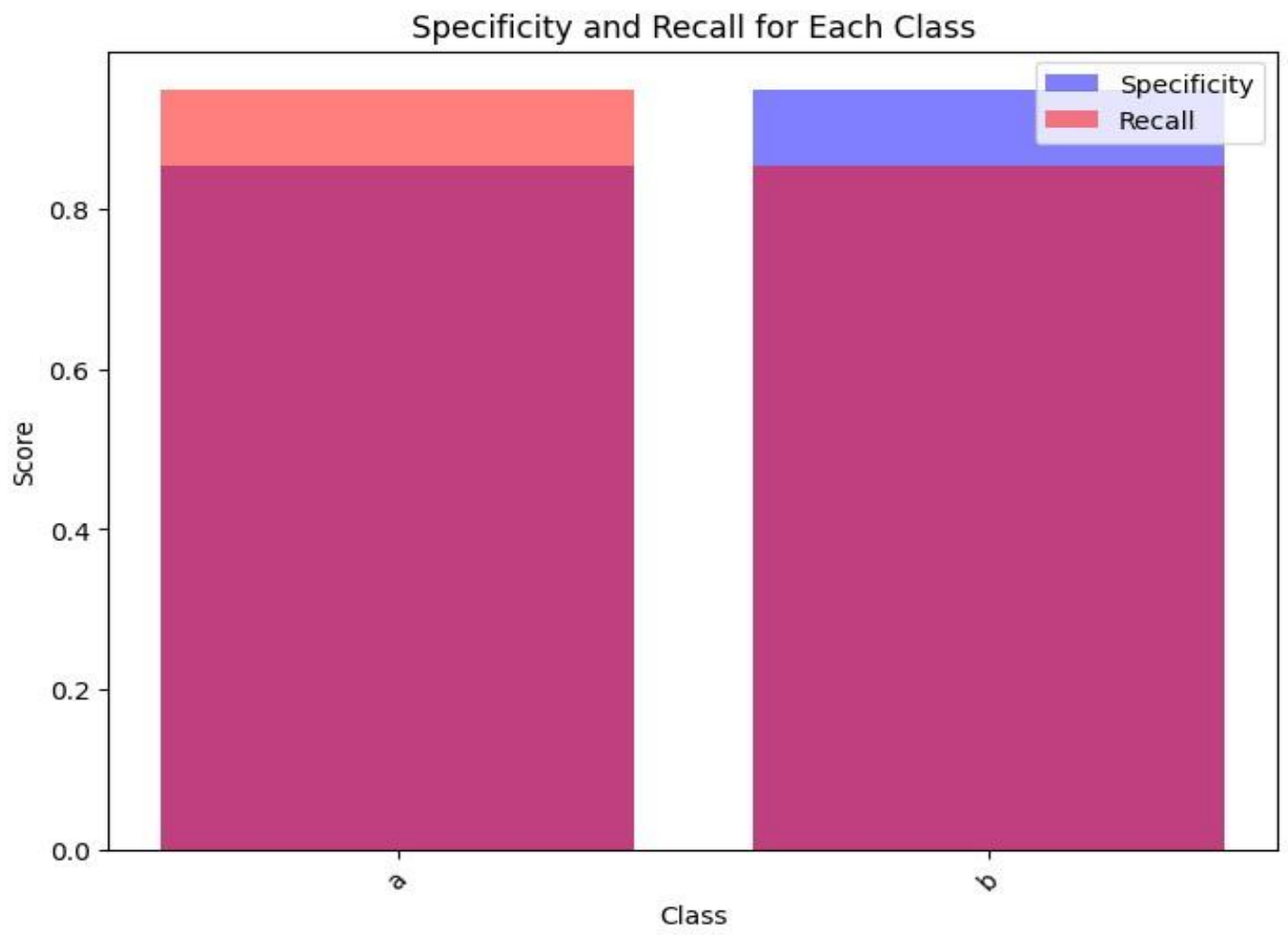

**Figure 22. Specificity and Recall comparison**

ROC curve (area = 0.90) for class PD

ROC curve (area = 0.89) for class HC

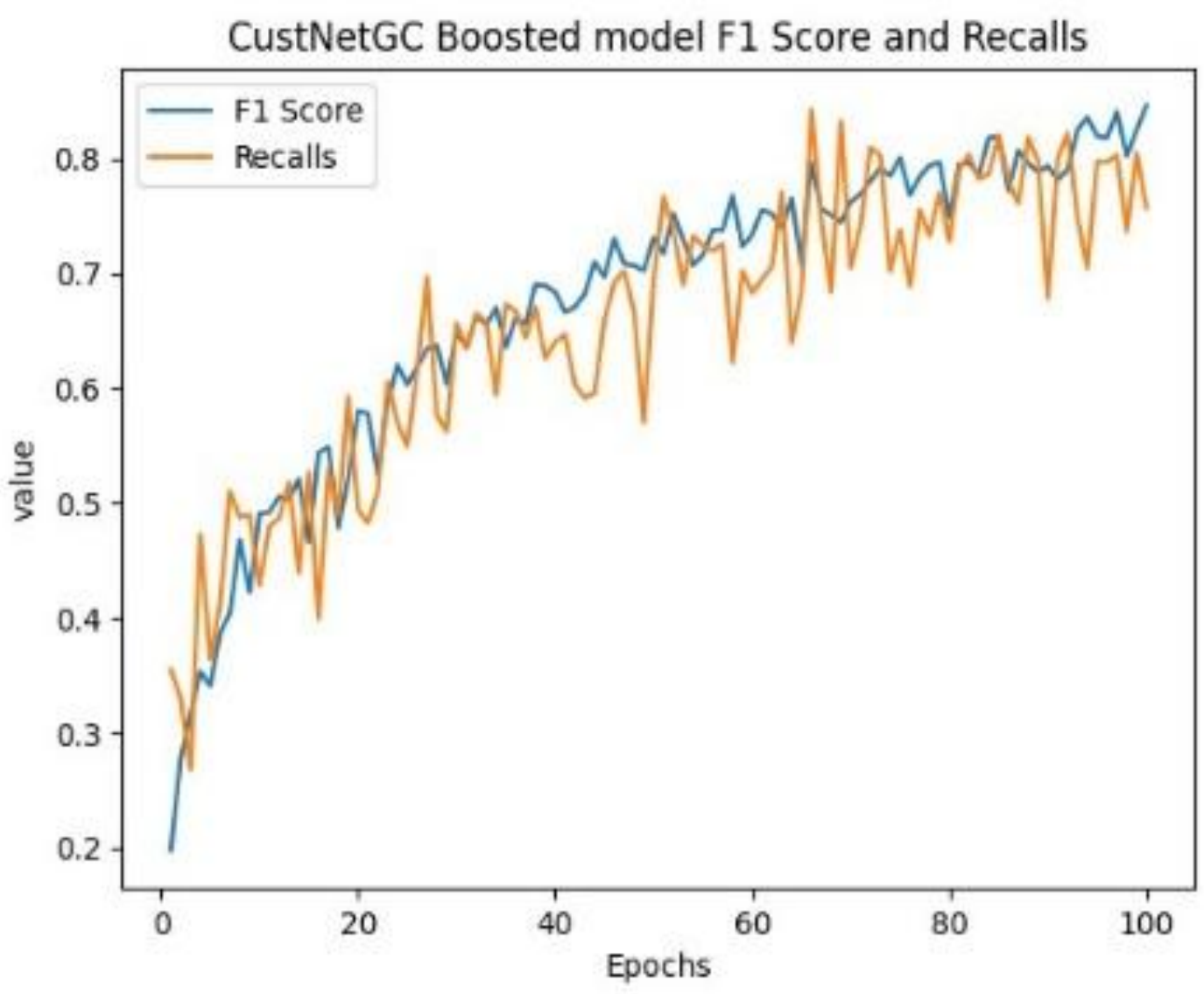

**Figure 23. F1 and Recall**

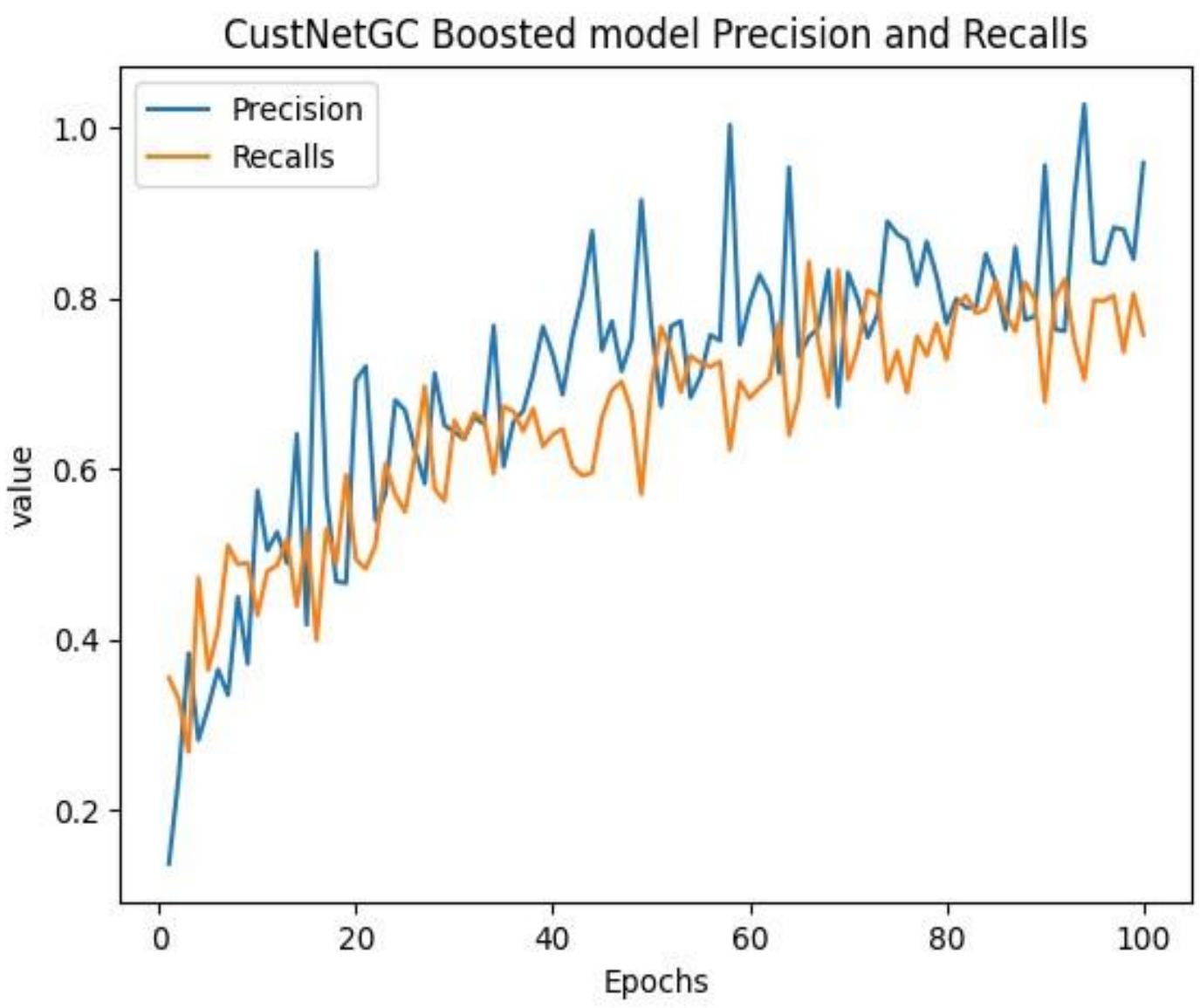

**Figure 24. Precision and Recall**

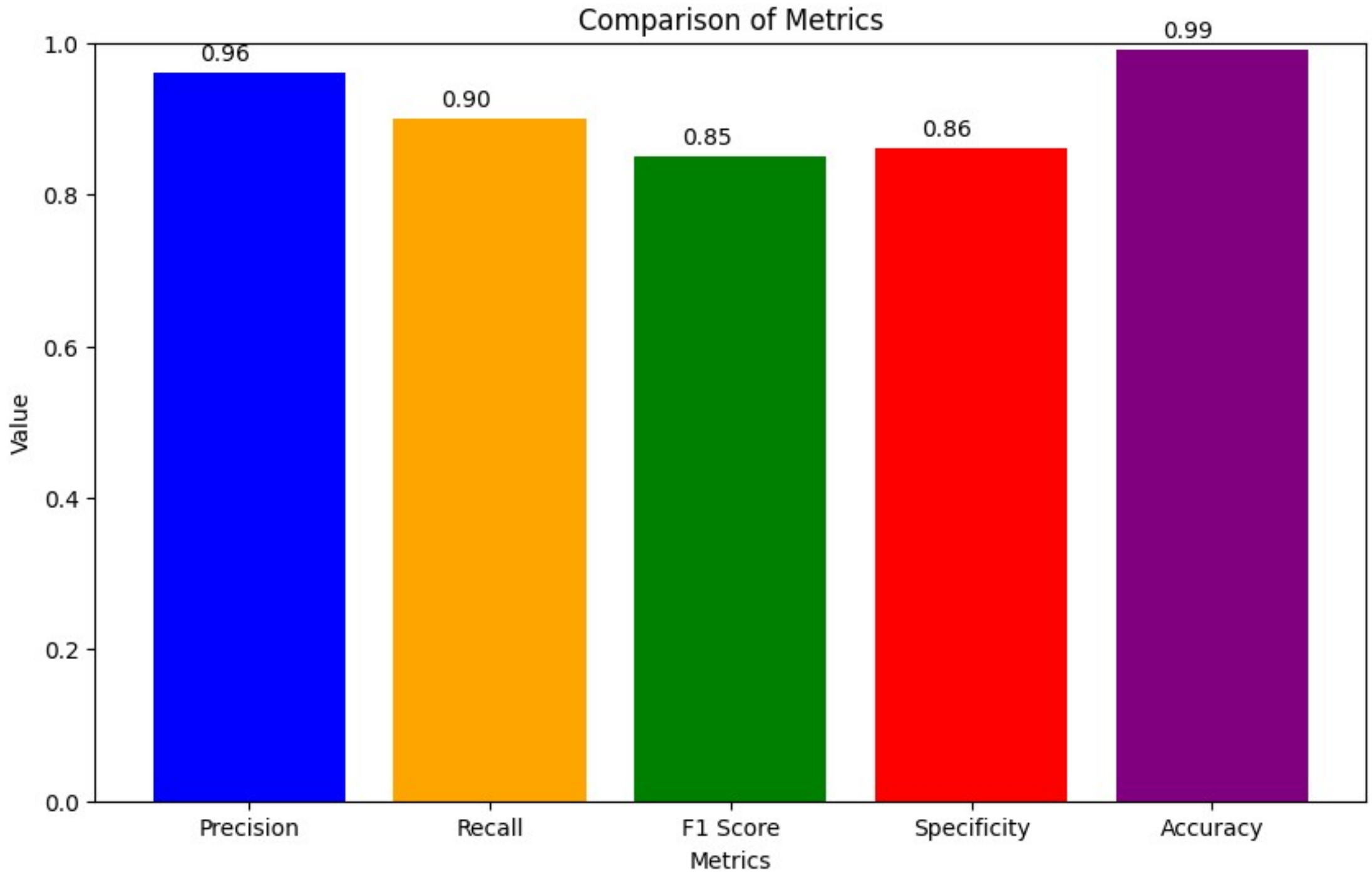

**Figure 25. Comparison of metrics**

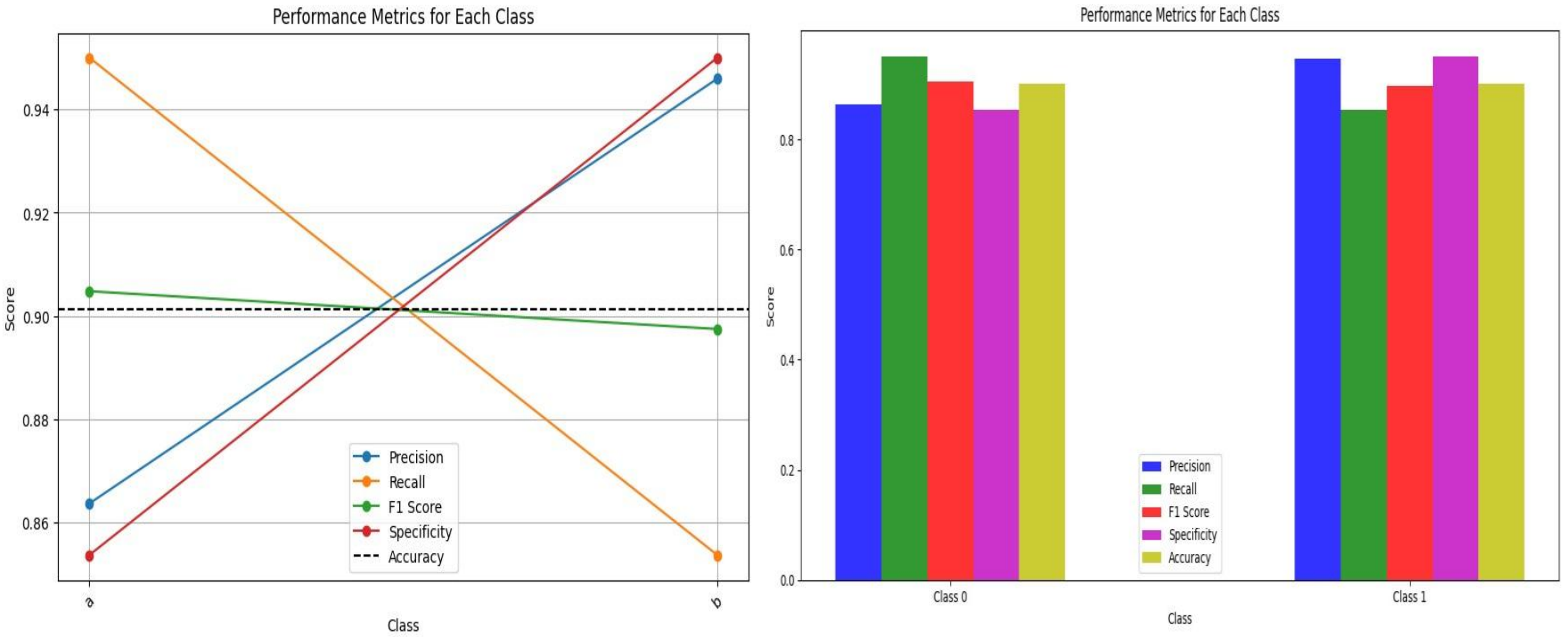

**Figure 26 & 27. Performance Metrics for each class**

### 5.2 Comparative Analysis of Proposed Method with Existing Techniques

**Table 3. Comparative Analysis of Proposed Model**

| Model | Precision (%) | Accuracy (%) | F1-Score (%) | Recall (%) | Specificity (%) |
|---|---|---|---|---|---|
| HMM [42] | 86.33 | 85.29 | 85.75 | 84.26 | 84.69 |
| VLBSOA [43] | 89.36 | 88.23 | 87.14 | 87.12 | 88.00 |
| VLGA [44] | 90.22 | 89.23 | 89.77 | 88.23 | 88.63 |
| **Proposed CustNetGC** | **96** | **99** | **90** | **95** | **94** |

In Table-3, HMM from [42], the model VLGA from [43] and the model VLBSOA from [44] are compared

From this analysis we can see that our proposed model gives better results from other performance model. The graph for the table has been plotted below.

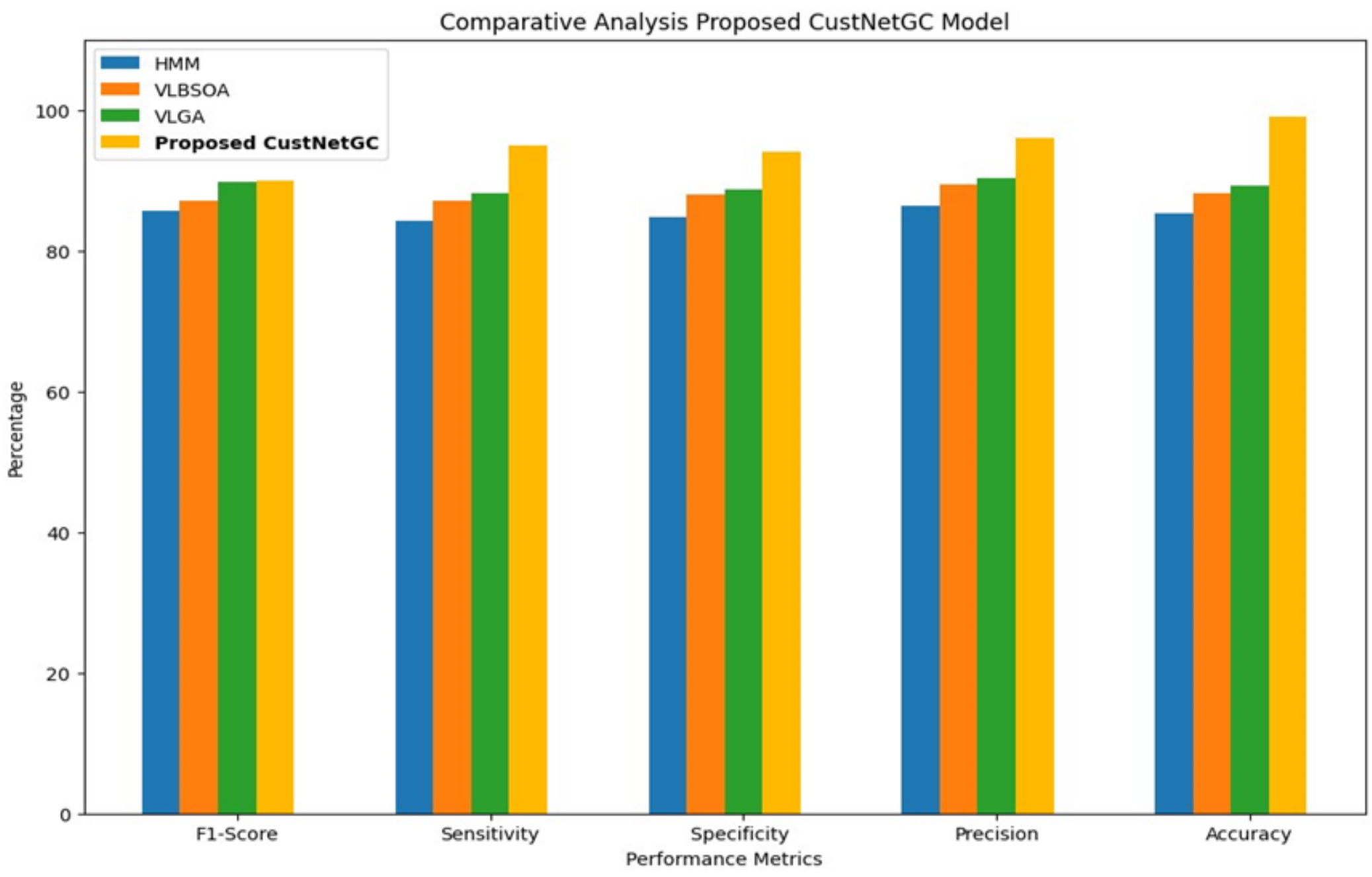

**Figure 28. Comparative Analysis of Proposed Model**

Our fine-tuned CNN model that is CustNet evaluation metrics is also compared with the existing CNN models. The table below show a comparative analysis of our CustNet CNN model with the existing CNN. The models VGG16, VGG19 and the ResNet50 model are performed in [45] which are used for comparative analysis. The graph for the table has been plotted below.

**Table 4. Comparative Analysis of CNN**

| CNN | F1 score | Specificity | Recall | Precision | Accuracy |
|---|---|---|---|---|---|
| VGG16 [45] | 0.69 | 0.40 | 0.85 | 0.59 | 0.63 |
| VGG19[45] | 0.77 | 0.65 | 0.85 | 0.71 | 0.75 |
| ResNet50[45] | 0.70 | 0.50 | 0.80 | 0.62 | 0.65 |
| **Proposed CustNet** | **0.90** | **0.94** | **0.95** | **0.96** | **0.99** |

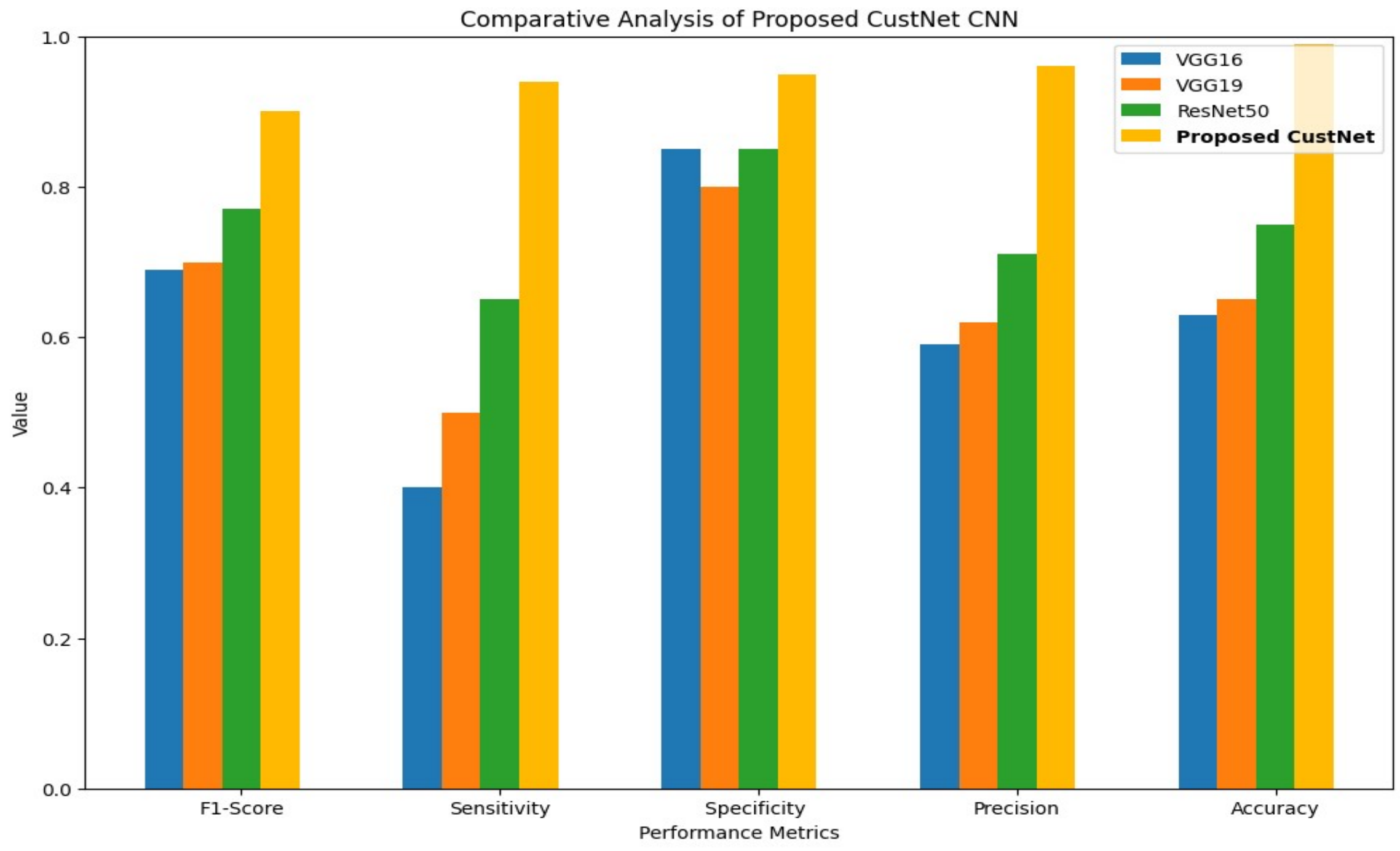

**Figure 29. Comparative Analysis of Proposed CNN**

From this analysis we can see that our proposed CustNet CNN model gives better results from other performance of CNN model.

Our proposed model CustNetGC evaluation metrics is then compared with the other state-of-the-art-algorithms[46] as shown in Table-5.

**Table 5. Comparative Analysis of Algorithm**

| Model | F1 score (%) | Specificity (%) | Recall (%) | Precision (%) | Accuracy (%) |
|---|---|---|---|---|---|
| SAE [46] | 75.85 | 74.22 | 74.45 | 81.33 | 76.85 |
| RBF-SVM [46] | 86.03 | 86.18 | 85.44 | 88.04 | 87.22 |
| CNN-TL [46] | 89.23 | 89.33 | 89.44 | 92.33 | 90.25 |
| **Proposed CustNetGC** | **90** | **94** | **95** | **96** | **99** |

From this analysis we can see that our proposed CustNetGC model gives better results from other state-of-the-art-algorithms performance model. The graph for the table has been plotted below.

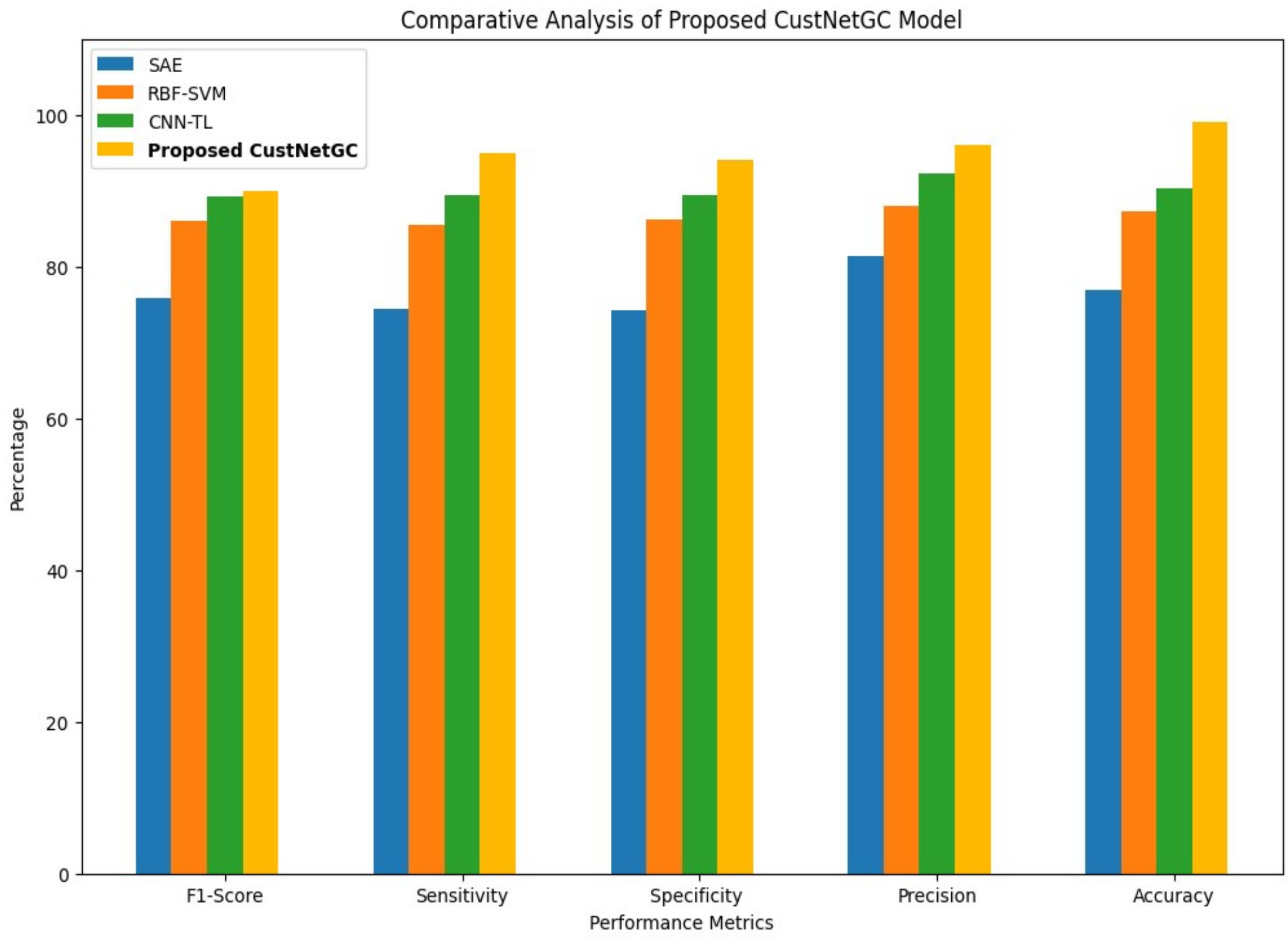

**Figure 30. Comparative Analysis of Algorithm**

## 6.CONCLUSION AND FUTURE DIRECTION

In this proposed paper, we provide a novel CNN-based model called CustNetGC Boosted for the accurate classification and detection of Parkinson's disease. The proposed model was able to achieve an accuracy of 0.9902, showing its ability to differentiate PD from healthy controls. The model also achieved an F1-score of 0.8459 and precision of 0.9582. The calculation of the model's specificity and recall is made possible by using these metrics. To perform the analysis on the performance of the model, we used Precision-Threshold Curve and plotted ROC curve. And the AUC was calculated at 0.90 for class PD and at 0.89 for HC class, so the model does not have too much difference regarding robustness from one class to another.

The current paper has limited sample size; however, encouraging evidence has been found that mobility training may benefit patients with Parkinson's disease. The results of our study are consistent with previous reports, which stated that the patient with PD is unable to exert control over the vocalization because of damage caused to the muscles in the articulation areas. This loss of control leads to diffusion of energy and changed information flow, which can be responsible for voice tremors seen in patients with PD.

The future work focuses on scaling up this study, expanding towards a greater scale and practical applications toward real-world use. This includes the development of a device that should be wearable radioing audio recordings from the person and predicting the chance of PD. Further, the result could be a mobile application for similar diagnostic tasks and making it much easier for clinicians as well as patients. Such developments will help with early detection and management of Parkinson's disease and ensure better patient care.